\documentclass[11pt]{article}

\usepackage[final]{acl}

\usepackage{times}
\usepackage{latexsym}

\usepackage[T1]{fontenc}

\usepackage[utf8]{inputenc}

\usepackage{microtype}

\usepackage{inconsolata}

\usepackage{graphicx}

\usepackage{enumitem}
\usepackage{amsmath}
\usepackage{amssymb}
\usepackage{booktabs}
\usepackage{array}
\usepackage{tabularx}
\usepackage{longtable}
\usepackage{multirow}
\usepackage[skins,breakable]{tcolorbox}
\usepackage{subcaption}

\newcolumntype{Y}[1]{>{\hsize=#1\hsize}X}

\setlist[itemize]{
    topsep=0pt,
    itemsep=0pt,
    parsep=0pt,
    partopsep=0pt,
    leftmargin=*
}

\newcommand{\CodeDataURL}{\url{https://github.com/SleepyWithoutCoffee/CATJudge}}

\title{Framework and Benchmark for Code-Driven Agentic Testing in Web Development}

\author{Bin Hong$^1$, Zhenchao Zhang$^3$, Jiyuan He$^3$, Kai Zhang$^1$, Zhenya Huang$^{1,2,\dag}$ \\
$^1$ State Key Laboratory of Cognitive Intelligence, \\ University of Science and Technology of China \\
$^2$ Institute of Artificial Intelligence, \\ Hefei Comprehensive National Science Center \\
$^3$ Meituan \\
\texttt{hb2002@mail.ustc.edu.cn}, \texttt{huangzhy@ustc.edu.cn}
}

\begin{document}
\maketitle
\def\thefootnote{\dag}\footnotetext{Corresponding author}\def\thefootnote{\arabic{footnote}}
\begin{abstract}
End-to-end GUI testing is essential for verifying web applications, yet existing evaluations rely on predefined checklists and are confined to the data and frameworks of web generation benchmarks, leaving the bug-discovery ability of vision-language models (VLMs) systematically untested. We introduce \textbf{C}ode-driven \textbf{A}gentic \textbf{T}esting (CAT), a paradigm in which the agent writes Playwright code to drive the browser, gathers feedback, and autonomously explores web applications to uncover bugs. We instantiate CAT with CATJudge, an agentic framework that unifies Browser-Use and Computer-Use tools within a single environment and CATTest, a benchmark of 102 AI-generated web applications with carefully annotated bugs, built through close human-AI collaboration to feature complex interactions and subtle defects. Experiments with mainstream VLMs show that all evaluated models perform poorly, revealing a clear gap between current VLM capabilities and the demands of real-world testing in AI web development. We release our code and data.\footnote{\CodeDataURL}
\end{abstract}

\section{Introduction}

With the rapid advancement of large language models (LLMs) in code intelligence, LLM-powered coding agents have achieved remarkable progress on repository-level code generation tasks~\citep{jimenez2024swe, jiang2026survey}. Collaborating with coding agents is increasingly becoming a programming paradigm that tangibly improves productivity~\citep{ge2025survey, jiang2024cursorcore}. While coding agents excel at conventional development, they still exhibit notable shortcomings in web application generation, which cannot be verified through textual unit tests or static analysis alone. Verifying web applications relies on visual and interactive feedback, as many bugs manifest only when the code is rendered in a browser~\citep{zhou2025janus, visconti2022webmonitor}. This makes end-to-end (E2E) testing involving graphical user interface (GUI) operations an essential measure for both quality assurance of AI-generated web applications and filtering of training data~\citep{ye2025ai, pezze2018automatic}.

E2E GUI testing incurs substantial human labor costs~\citep{alegroth2015visual, pezze2018automatic}. Benchmarks on web code generation typically employ vision-based rule scoring or LLM-as-Judge scoring~\citep{sun2025fullfront, wang2025webgen, awal2025webmmu}. More recently, benchmarking efforts have focused on using GUI agents based on vision-language models (VLMs) to automatically execute encapsulated GUI operations such as clicks and type to verify web code correctness~\citep{lu2026webgen, wu2025frontalk, bian2025you}. However, current LLMs/VLMs have limited GUI capabilities, and their GUI interaction environments typically assume fully functional web pages~\citep{xie2024osworld, miyai2025webchorearena}. For testing scenarios that require identifying bugs based on anomalous feedback and demand more intensive web development knowledge, there is a lack of systematic evaluation of their E2E GUI testing abilities.

This gap manifests in two aspects. First, the current web testing paradigm relies on predefined checklists (GUI test cases described in natural language)~\citep{lu2026webgen, wu2025frontalk}. Yet AI-generated web pages, even when constrained by prompts, often produce features beyond what the checklist anticipates. The testing pipeline can only reflect the model's ability to execute predefined GUI test cases. More importantly, as code intelligence continues to advance and AI-generated web pages grow in complexity, the volume of checklist items will inevitably expand, leading to higher costs, decreased testing efficiency, and reduced reliability. Second, current evaluations of GUI testing capabilities are merely validations of automated evaluation methods within web generation benchmarks, constrained by the data and frameworks of those benchmarks themselves. The types and complexity of interactions required by the web pages are limited, and the subtlety of bugs is also limited.

To bridge this gap, we first model the pipeline from user-provided prompts to model-delivered code in AI web generation scenarios. Based on this, we define a different testing paradigm: the agent directly writes Playwright~\citep{playwright2025} code to interact with the browser, gather textual and visual feedback, and autonomously explore the web application to discover bugs. We term this paradigm as \textbf{C}ode-driven \textbf{A}gentic \textbf{T}esting (CAT).
Then, we build CATJudge, an agentic framework that unifies the two mainstream families of GUI agent tools, Browser-Use Agent (BUA)~\citep{he2024webvoyager} and Computer-Use Agent (CUA)~\citep{xie2024osworld} tools, within a single environment, allowing both toolsets to be invoked for web exploration. To evaluate the CAT capabilities of VLMs under the CAT paradigm, we introduce CATTest, a benchmark of 102 AI-generated web applications and carefully annotated bugs constructed through iterative, close human-AI collaboration. Task instances in CATTest include complex interactions and subtle bugs, demanding strong Playwright coding ability and long-horizon abilities. 

We evaluate various VLMs on the CATTest dataset using CATJudge. Experimental results show that current mainstream VLMs achieve poor performance on the CATTest dataset (R-score below 43), revealing a substantial gap between the capabilities of existing VLMs and the practical requirements of real-world web development testing.

Our contributions are summarized as follows:
\begin{itemize}
    \item We develop a theoretical workflow model for systematically analyzing the agentic web generation scenario, guiding the construction of related framework and benchmark.
    \item We propose CATJudge, an agentic framework along with a unified environment for executing CAT tasks for web applications. CATJudge integrates both BUA and CUA tools, and executes actions in real browsers.
    \item We construct CATTest, an agentic benchmark of 102 complex web applications for evaluating the CAT capabilities of VLMs in web development.
    \item We conduct experiments on CATTest with multiple mainstream VLMs using CATJudge, providing insights into improving their CAT capabilities in web development.
\end{itemize}

\section{Related Works}

\paragraph{Code Generation Benchmarks}
As the coding capabilities of LLMs have ~\citep{ye2026themis, jiang2025verse, zhao2024repair}, research efforts in code intelligence have gradually shifted from function-level generation (e.g., HumanEval~\citep{chen2021evaluating} and LiveCodeBench~\citep{jain2025livecodebench}) toward file-level and repository-level tasks that align with real-world needs (e.g., SWE-Bench~\citep{jimenez2024swe} and Terminal-Bench~\citep{merrill2026terminal}). These benchmarks generally rely on pre-written static test scripts and test cases to verify the generated code. 
Web development is a common repository-level code generation scenario in real-world software engineering. Numerous benchmarks have been proposed to evaluate LLMs/VLMs on generating web code. In terms of task formulation, these benchmarks fall into two categories: (1) generating web code from visual information (e.g., images~\citep{sun2025fullfront, si2025design2code, awal2025webmmu} and videos~\citep{chen2025iwr}), and (2) generating web code from textual prompts~\citep{lu2026webgen, zhang2025artifactsbench, sun2025fullfront, wu2025frontalk}. To verify AI-generated web code, these benchmarks employ a variety of approaches, including rule-based validation, LLM-as-Judge scoring, and using GUI agents to simulate human interactions.


\paragraph{GUI Agents and Why GUI Agents}
GUI agents are a class of agents equipped with specialized interaction tools that receive visual information or accessibility information (e.g., accessibility tree, a11y tree) and interact with real environments through these tools~\citep{nguyen2025gui, hu2025agents}. They are primarily divided into two categories: Browser-Use Agents (BUA)~\citep{browser-use2025, he2024webvoyager} and Computer-Use Agents (CUA)~\citep{xie2024osworld, wang2025ui}. The former is equipped with tools built on browser automation frameworks (e.g., Playwright~\citep{playwright2025} and Selenium~\citep{selenium2025}), enabling LLMs to interact with browsers. The latter is equipped with system-level keyboard and mouse automation tools such as PyAutoGUI~\citep{pyautogui2025}, enabling LLMs to localize targets on the screen and interact with computers using mouse and keyboard input just as humans do. For scenarios such as web development involving visual understanding and physical interaction, end-to-end GUI testing is a critical method for finding bugs that appear correct at the textual code level yet manifest as errors when deployed in a browser. Rule-based validation and LLM-as-Judge scoring or comparison struggle to reflect the interactive functionality of web pages after actual deployment. Consequently, recent web generation benchmarks have increasingly adopted GUI agents to automate interaction with web pages for testing~\citep{lu2026webgen, chen2025iwr, wu2025frontalk}.


\section{CAT in Web Development}
\label{sec:catjudge}
In this section, we introduce \textbf{C}ode-Driven \textbf{A}gentic \textbf{T}esting in Web Development in detail. We begin with our research methodology, in which we model the workflow of AI web generation and define the CAT paradigm. Building on this, we design the CATJudge framework. Finally, we construct the CATTest benchmark for evaluation.

\subsection{Methodology}

\begin{figure}[t]
  \centering
  \includegraphics[width=0.98\columnwidth]{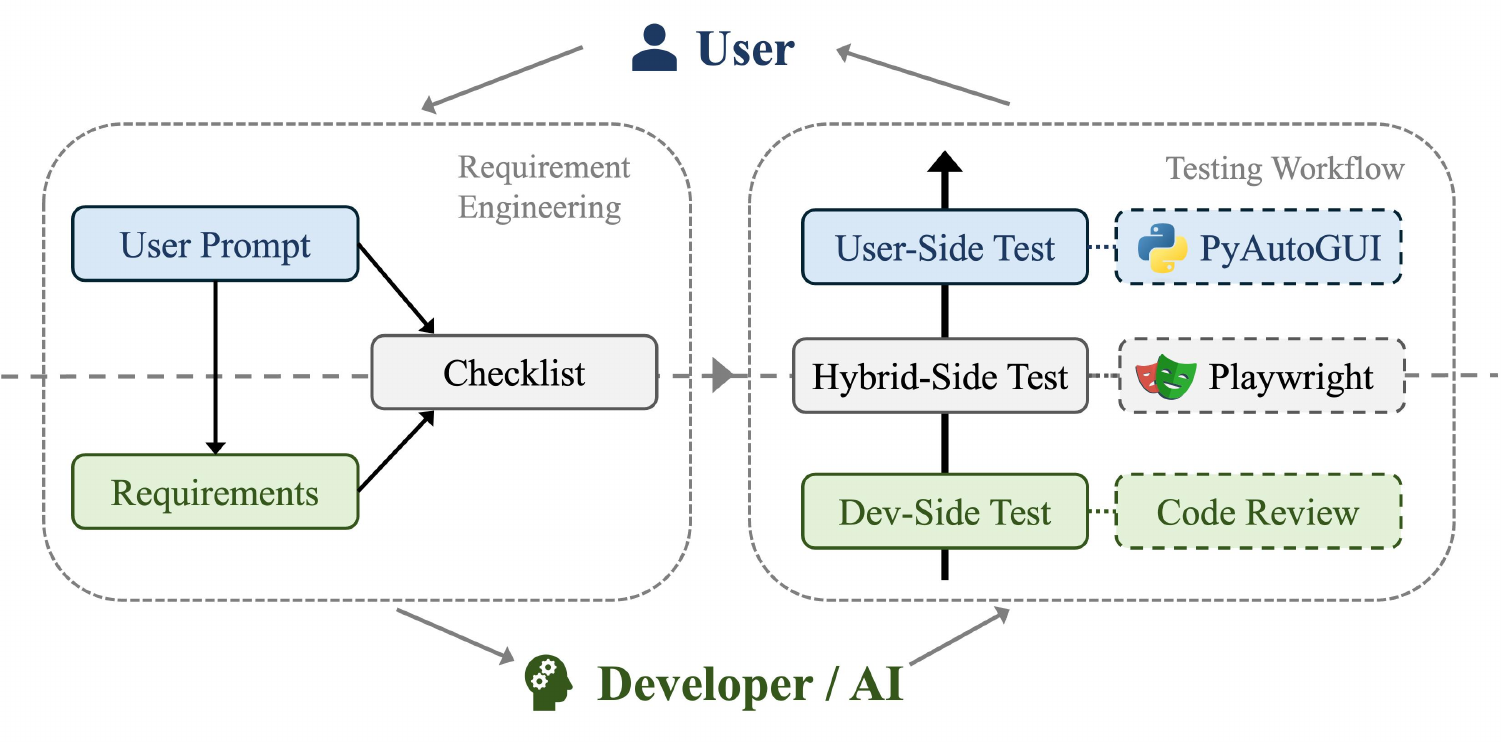}
  \caption{Abstracted workflow of AI-driven web development.
  Our evaluation focuses on the testing workflow.}
  \label{fig:workflow}
\end{figure}

Analogous to the human software development process, the AI-driven web development workflow can be abstracted as a closed loop: user $\rightarrow$ requirements $\rightarrow$ AI $\rightarrow$ testing $\rightarrow$ delivery, as shown in Figure~\ref{fig:workflow}.
In this model, the user provides a prompt, which is then interpreted to produce a requirements document. Developers implement requirements, while QA engineers design checklists. Unlike the human development scenario, AI generates code faster and therefore demands a more efficient workflow with less communication and faster testing. To achieve this goal, the link between requirement engineering and the testing workflow in Figure~\ref{fig:workflow} can be removed: the model needs to autonomously discover user-perceivable bugs through review or testing, without relying on a predefined checklist.

The testing workflow consumes information handed off directly from the development side and proceeds through layers of inspection. (1) \textit{Developer-Side}: an inspection that requires professional expertise and is conducted from the developer's perspective, such as code review. (2) \textit{Hybrid-Side}: an inspection that combines professional expertise with user-perspective experience, e.g., examining page source and element rendering with browser developer tools, such as Playwright-based testing. (3) \textit{User-Side Test}: an inspection accessible even to users without specialized expertise, who can only rely on keyboard/mouse interactions and visual feedback, such as PyAutoGUI-based testing. Purely developer-side inspection lacks interaction and visual feedback, whereas purely user-side testing is constrained by the model's on-screen element-localization ability and cannot leverage the assistance provided by the browser developer tools. We therefore evaluate the VLM's capability for Hybrid-Side testing with Playwright. The VLM writes Playwright code and uses it to interact with the browser, gather textual and visual feedback, and autonomously explore the web application to discover bugs. We term this paradigm as \textbf{C}ode-driven \textbf{A}gentic \textbf{T}esting (CAT).

\subsection{CATJudge Framework}

\begin{figure}[t]
  \centering
  \includegraphics[width=0.98\columnwidth]{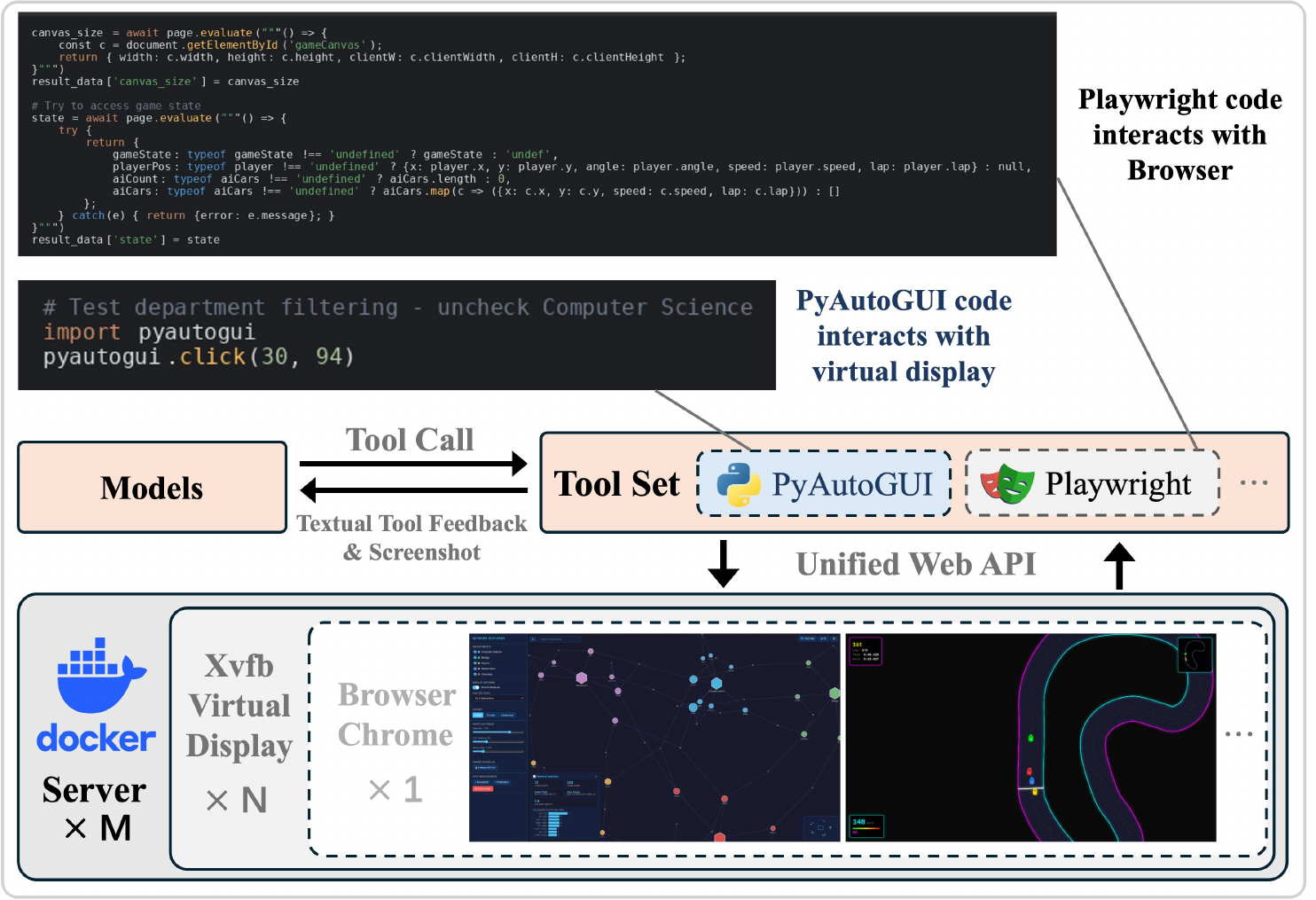}
  \caption{Overview of the CATJudge framework (set-up and tear-down procedures omitted).}
  \label{fig:catjudge}
\end{figure}
\paragraph{Task Definition}
We formulate autonomous CAT as a POMDP $(\mathcal{S}, \mathcal{A}, \mathcal{O}, \mathcal{T}, \Omega, \mathcal{R}, H)$ with horizon $H$. A state $s_t \in \mathcal{S}$ is the latent environment state of the deployed web application (e.g., DOM and runtime state, together with its unknown defects). At each turn the agent receives an observation $o_t \sim \Omega(\cdot \mid s_t)$ exposing only part of $s_t$ through its tools (\S\ref{sec:workflow}), and selects a tool action conditioned on the task context $\mathcal{I}$ (the testing instruction together with a QA-oriented project document, see \S\ref{sec:workflow}) and the history $h_t = (\mathcal{I}, o_{0:t}, a_{0:t-1})$.

To keep the toolkit GUI-only, the action space contains only interaction tools and a termination tool $\mathcal{A} = \mathcal{A}_{\text{int}} \sqcup \{a_{\text{term}}\}$: interaction tools in $\mathcal{A}_{\text{int}}$ drive $s_t$ via $\mathcal{T}$, while the termination action $a_{\text{term}}$ ends the episode and emits a verdict (\textsc{pass} or \textsc{not\_pass}) together with a textual list of discovered bugs $\mathcal{B}$. The episode also ends when $H$ turns are reached. Letting $T$ denote the terminal turn, we adopt a sparse, episodic reward
\begin{equation}
\mathcal{R}(s_T, a_T) = \mathrm{score}(\mathcal{B},\, \mathcal{B}^{*}) \in [0, 1],
\end{equation}
where $\mathcal{B}^{*}$ is the ground-truth bug set and $\mathrm{score}(\cdot,\cdot)$ is defined in \S\ref{eval}.

\subsubsection{Environment}
\label{sec:environment}

\paragraph{Remote Testing Service} For extensibility and flexible deployment, CATJudge runs the web application under test on a remote server (e.g., a Docker container) that is decoupled from the local client. Before each testing episode, the server automatically executes a predefined set of deployment commands (e.g., \texttt{npm install} and \texttt{npm run dev}) to bring up the application, and tears the service down upon episode termination or interruption. The execution interfaces of all tools are exposed to the local tool interpreter through a unified web API, so the agent can invoke remote tools through the client as if they were local.

\paragraph{Hybrid Automation \& Task Isolation} 
The architectures of existing BUA and CUA differ from each other~\citep{he2024webvoyager, browser-use2025, xie2024osworld}, and there is a need for a unified runtime framework.
As shown in Figure~\ref{fig:catjudge}, to accommodate both browser-only BUA tools via Playwright and OS-level CUA tools for keyboard/mouse automation via PyAutoGUI within a single environment, the server launches an Xvfb virtual display: CUA actions can run on this virtual screen, while the browser is started and rendered on the same virtual display so that BUA actions can be executed in a real browser. This co-located design allows CUA and GUI automation to coexist and to interact with a consistent rendering of the web page. Building on this design, the framework supports two complementary isolation modes: run actions on different servers yields strong isolation, whereas allocating different virtual displays on a single server provides lightweight, soft isolation.

\subsubsection{Workflow}
\label{sec:workflow}
\paragraph{Initial State} Recent studies suggest that LLMs exhibit clear limitations in exploring and learning unfamiliar environments from scratch~\citep{foundation2026arc}. To improve testing efficiency, we therefore inject a QA-oriented project document into the initial state. As is common practice in AI-assisted coding pipelines, the document is supplied by the development side and offers a minimal yet sufficient guide for testing: an overview of implemented features, the technical context (e.g., tech stack, core logic, and key state variables), and the interaction surface relevant to testing. We also provide an initial screenshot of the web page.

\paragraph{Feedback \& Observation} Each tool call returns textual feedback that aggregates the kinds of signals a human QA engineer would obtain from invoking the corresponding tool during testing. For example, a Playwright call may return execution status and error messages, browser state and return values from injected scripts and DevTools commands. After all tool calls in a turn have executed, the server waits for a short delay to let the page settle in response to the interactions, and then captures a screenshot of the resulting state and returns it to the agent as part of the observations. To bound the context length, only the most recent screenshot is retained in the history, which also matches the visual feedback a human tester would actually have access to at any moment.

\subsubsection{Tools}
\label{sec:tools}

\paragraph{GUI Tools} Our framework exposes BUA and CUA tools for GUI interactions. Each contains: (1) a set of encapsulated atomic operations following prior works~\citep{he2024webvoyager, xie2024osworld}; and (2) a script-execution interface through which the agent can freely author and run code.

\paragraph{Core Tool Set} Our experiments show that code-capable models prefer the more flexible script-execution tool over the atomic operations (see \S~\ref{sec:analysis}). We argue that, for LLMs, interacting with the environment by writing code is more native and more efficient than imitating low-level human action sequences. We therefore include only the script-execution tool into the core tool set for GUI interactions. In addition, we provide a special tool for emitting the termination action. Departing from prior work that feeds the a11y tree to the model as part of the input~\citep{he2024webvoyager}, we expose the a11y tree as a standalone tool that the agent may invoke at will, while screenshots are instead pushed to the agent at every turn to mitigate the mirage reasoning~\citep{asadi2026mirage} issue. We also provide a wait tool that lets the agent intentionally idle, either to allow time-dependent operations to take effect or to actively request a fresh screenshot.

\begin{figure}[t]
  \centering
  \includegraphics[width=0.98\columnwidth]{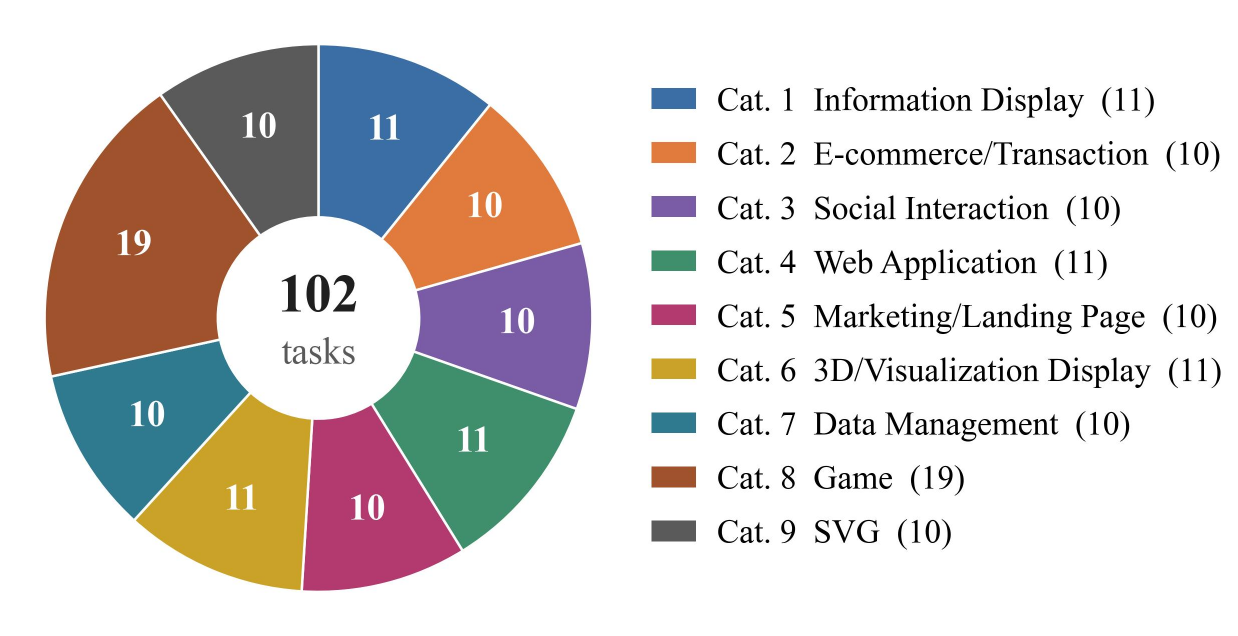}
  \caption{Distribution of the $102$ tasks in CATTest across the $9$ top-level categories.}
  \label{fig:category_taxonomy}
\end{figure}

\subsection{CATTest Benchmark}
\label{sec:cattest}
To rigorously evaluate the code-driven agentic testing capability of VLMs, we construct CATTest, a benchmark dataset consisting of complex AI-generated web applications, paired with project documentations and carefully annotated ground-truth bugs (GT bugs).

\subsubsection{Benchmark Construction}
\label{sec:benchmark_construction}
\begin{figure*}[t]
  \centering
  \includegraphics[width=0.97\textwidth]{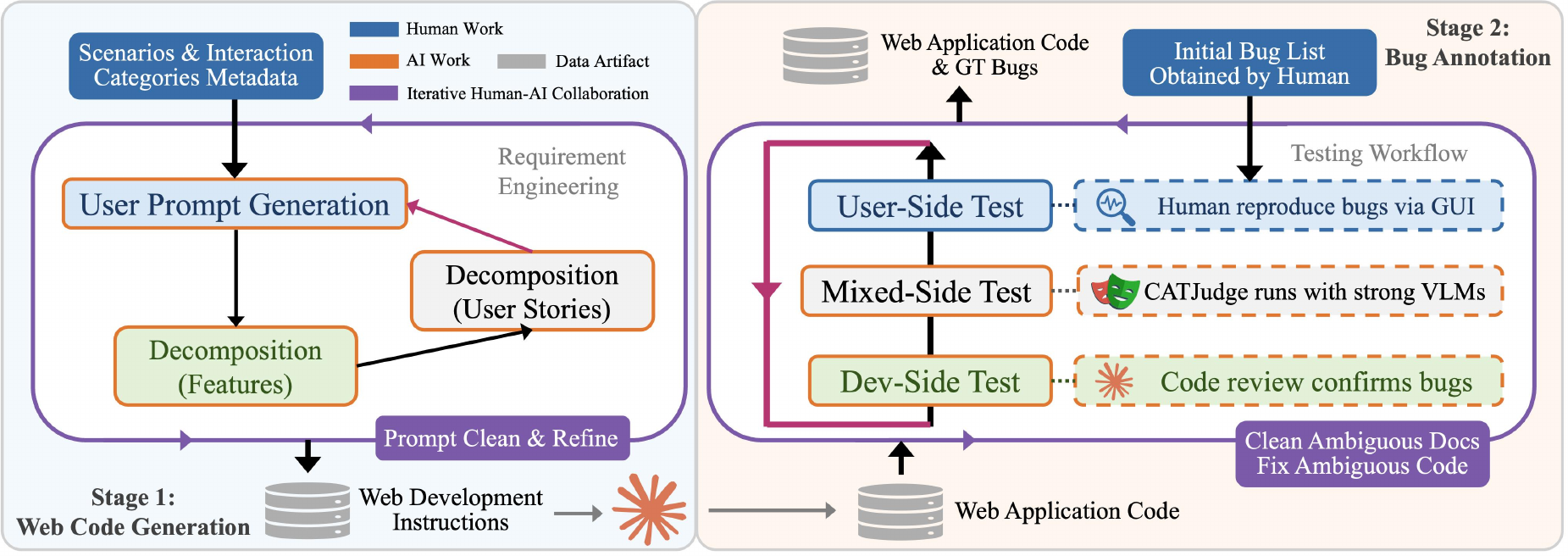}
  \caption{Overview of the CATTest benchmark construction. Dark blue denotes human work steps, orange denotes AI work steps, purple denotes iterative human-AI collaborative steps, and grey denotes data artifacts.}
  \label{fig:benchmark_construction}
\end{figure*}

The construction of CATTest is a process of close human-AI collaboration, whose overall pipeline is illustrated in Figure~\ref{fig:benchmark_construction}. In total, CATTest contains 102 data points. Detailed statistics of CATTest are reported in Appendix~\ref{app:statistics}.

\paragraph{Web Code Generation} To ensure broad coverage of interaction types, we first identify 9 popular usage scenarios for web applications. Within each scenario, drawing on representative real-world use cases and on category definitions on Wikipedia, we enumerate the interactions (e.g., element dragging, mouse-driven viewport rotation) required by each application, while keeping the overlap of required interactions among applications within the same top-level category as small as possible. This yields a batch of category metadata.

To ensure that GT bugs are sufficiently subtle and natrually, the generation tasks must be hard enough on one hand, and the code generator must be capable enough on the other, so that the elicited bugs reflect real defects that genuinely occur in AI coding workflows. To raise the difficulty of the generation tasks, we iteratively expand the requirement volume of each prompt following the requirement engineering workflow in Figure~\ref{fig:workflow}. Specifically, we first attach a concrete fictional usage context (e.g., an advertising site for a coffee product) to each task, and use the metadata to prompt Claude-Opus-4.6~\citep{anthropic2026claudeopus46} for initial seed web development instructions. Following standard agile-development practice~\citep{manifesto2001manifesto}, we then perform a two-step decomposition of the seed instructions into feature lists that reflect coarse-grained requirements and user stories that reflect user paths. At each iteration, we leverage the user stories from the previous round to enrich the prompt with finer details and additional features, and redo the decomposition until the resulting user stories are sufficiently numerous, which in turn guarantees the difficulty of the task. The instructions are subsequently reviewed and cleaned by human annotators: (1) check whether the instruction targets a real-world website, and remove any related content to avoid copyright concerns; (2) strip over-specific technical constraints (e.g., self-containment requirements) so that the technology usage and distribution more closely match real-world AI coding scenarios; (3) remove requirements that CATJudge cannot support, e.g., replacing a real shopping transaction flow with a mock to avoid third-party services, or simulating a gyroscope with the mouse on a desktop environment; and (4) simplify tasks whose content volume exceeds the capacity of any available AI tool, reducing irrelevant textual content and repetitive technical points while preserving interaction richness (e.g., for a page that showcases product manuals, reducing the amount of product description text and the number of chapters).

To maximize the chance that bugs genuinely populate the coding-agent regime, we generate the web pages with Claude Code~\citep{anthropic2025claudecode} backed by Claude-Opus-4.6, which represents the frontier of coding-agent capability, and obtain the corresponding project documents along the way. Appendix~\ref{app:generator_choice} discusses this choice in detail. Finally, human annotators experience each generated web application to confirm the technical feasibility of the development instruction as well as the complexity and interactivity of the resulting project.

\paragraph{Bug Annotation} This step is the most labor-intensive and the most challenging stage of the construction pipeline, as no single tool or expert can uncover every bug~\citep{su2021benchmarking}. After manual testing, CATJudge can still discover a substantial number of bugs that human testers miss. We therefore undertake a heavy iterative pipeline of annotation and verification, following the testing workflow in Figure~\ref{fig:workflow}. First, the human experience step in data construction yields an initial bug list. For each bug on the list, a human annotator attempts to reproduce it through manual GUI actions to confirm the described phenomenon truly occurs. Claude Code backed by Claude-Opus-4.6 then analyzes the source code and the project document with respect to the bug report to determine whether the source code is indeed defective, with humans collaboratively working with it throughout the process. To reduce subjective judgment during evaluation and improve its stability, ambiguous wording in the generation instruction and the project document are simultaneously cleaned and rewritten to handle false positives caused by unclear requirements or misleading documentation (e.g., the drag-to-zoom direction is counter-intuitive). For bugs remaining ambiguous (e.g., pixel-level minor UI offsets), we instead modify the source code directly to eliminate them. To maximize bug discovery coverage, we adopt a tool-competition pooling strategy~\citep{metzman2021fuzzbench}: run CATJudge with multiple strong code-capable models (see Appendix~\ref{app:model_in_annotation} for details), contribute additional bugs to the list, after which the bug confirming procedure is repeated. We iterate the loop until human annotators and Claude Code agree that every newly reported bug is either a false positive or a duplicate of a confirmed one (describing the same phenomenon, or textually being a cause of another). This yields the final GT bug set for evaluation.

\begin{table*}[t]
  \centering
  \small
  \begin{tabularx}{0.98\textwidth}{l *{6}{>{\centering\arraybackslash}X} @{}}
    \toprule
    \textbf{Model} & \textbf{R-score}$\uparrow$ & \textbf{P-score}$\uparrow$ & \textbf{Acc}$\uparrow$ & \textbf{Recall}$\uparrow$ & \textbf{Precision}$\uparrow$ & \textbf{F1}$\uparrow$ \\
    \midrule
    \multicolumn{7}{@{}c}{\textbf{Proprietary Models}} \\
    \midrule
    Claude-Opus-4.7        & \textbf{42.57} & 38.30 & \textbf{83.33} & \underline{93.67} & 86.05 & \textbf{89.70} \\
    Claude-Opus-4.6        & \underline{41.55} & \underline{48.20} & 81.37 & 89.87 & 86.59 & 88.20 \\
    Claude-Sonnet-4.6      & 37.68 & 41.08 & \underline{82.47} & \textbf{98.65} & 82.02 & \underline{89.57} \\
    Gemini-3.1-Pro         & 36.49 & \textbf{48.56} & 73.53 & 82.28 & 83.33 & 82.80 \\
    Gemini-3-Flash         & 24.22 & 22.99 & 73.53 & 83.54 & 82.50 & 83.02 \\
    Gemini-3.1-Flash-Lite  & 20.92 & 23.37 & 48.04 & 40.51 & 84.21 & 54.70 \\
    GPT-5.5                & 39.41 & 46.88 & 76.47 & 86.08 & 83.95 & 85.00 \\
    GPT-5.4                & 17.58 & 22.37 & 76.47 & \underline{93.67} & 79.57 & 86.05 \\
    GPT-5.4-mini           & 19.07 & 25.00 & 58.82 & 62.03 & 80.33 & 70.00 \\
    GLM-5V-Turbo           & 24.67 & 33.50 & 61.78 & 58.23 & \underline{86.79} & 69.70 \\
    Doubao-Seed-2.0-Pro    & 22.79 & 26.55 & 50.98 & 44.30 & 85.37 & 58.33 \\
    Doubao-Seed-2.0-Lite   & 21.98 & 23.20 & 42.16 & 32.91 & 81.25 & 46.85 \\
    \midrule
    \multicolumn{7}{@{}c}{\textbf{Open Models}} \\
    \midrule
    Qwen3.5-Plus           & 24.71 & 32.62 & 68.63 & 72.15 & 85.07 & 78.08 \\
    Qwen3.6-27B            & 33.56 & 37.92 & 71.57 & 72.15 & \textbf{89.06} & 79.72 \\
    Qwen3.5-27B            & 23.24 & 26.72 & 69.61 & 78.48 & 81.58 & 80.00 \\
    Qwen3.5-9B             & 14.71 & 14.71 & 43.14 & 25.37 & 68.00 & 36.96 \\
    Kimi-K2.6              & 30.78 & 29.43 & 69.61 & 75.95 & 83.33 & 79.47 \\
    \bottomrule
  \end{tabularx}
  \caption{Main results of mainstream VLMs on CATTest under the CATJudge framework. R-score and P-score are the fine-grained metrics measuring bug-discovery ability. Acc, Recall, Precision, and F1 are the coarse-grained pass/not-pass classification metrics. \textbf{Bold} marks the best result per column, and \underline{underline} marks the second best.}
  \label{tab:main_results}
\end{table*}

\subsubsection{Evaluation Metrics}
\label{eval}
To ensure that the evaluation metrics faithfully reflect the practical needs of AI coding, we design two complementary sets of metrics, at the coarse-grained and fine-grained levels respectively.

\paragraph{Coarse-grained Metrics} We treat the verdict of whether a project passes testing as a binary classification task. Taking \textsc{not\_pass} as the positive class, we report the four classical metrics: Accuracy, Recall, Precision, and F1. Coarse-grained metrics do not fully reflect a model's true bug-finding ability, but they can serve as a reference for filtering training data for web code generation tasks.

\paragraph{Fine-grained Metrics} For each project, we assign a score in $[0, 1]$ based on how many of the GT bugs are reported. We first use Gemini-3-Flash~\citep{google2025gemini3flash} to match the model reported bugs $\mathcal{B}$ against the GT bugs $\mathcal{B}^{*}$. The matching criterion is the same as in annotation: two bugs are considered matched if they describe the same phenomenon, or if one is a cause or a direct consequence of the other. The full criterion, the matching procedure, and worked examples are given in Appendix~\ref{app:bug_matching}. We performed manual bug matching on the test outputs of Claude-Opus-4.7 and compared it with the automatic matching; as shown in Table~\ref{tab:bug_matching} (\S\ref{sec:analysis}), the automatic matching closely aligns with human judgment. Let $M \subseteq \mathcal{B} \times \mathcal{B}^{*}$ denote the set of matched pairs
  produced by the matcher, and write
  $\mathcal{B}^{*}_{M} = \{b^{*} \in \mathcal{B}^{*} \mid \exists b,\, (b, b^{*}) \in M\}$
  and $\mathcal{B}_{M} = \{b \in \mathcal{B} \mid \exists b^{*},\, (b, b^{*}) \in M\}$.
We compute the scores as follows:
\begin{itemize}
    \item \textbf{Recall score (R-score)}, the primary metric of CATTest, which measures how comprehensively the model uncovers bugs. A higher score indicates that more GT bugs are caught:
    \begin{equation}
    R =
    \begin{cases}
    1, & |\mathcal{B}^{*}| = 0 \text{ and } |\mathcal{B}| = 0, \\
    0, & |\mathcal{B}^{*}| = 0 \text{ and } |\mathcal{B}| > 0, \\
    \dfrac{|\mathcal{B}^{*}_{M}|}{|\mathcal{B}^{*}|}, & |\mathcal{B}^{*}| > 0,
    \end{cases}
    \end{equation}

    \item \textbf{Precision score (P-score)}. Since the R-score does not penalize over-reporting, we additionally compute a P-score that measures the model's tendency to report bugs. A higher score indicates more accurate reporting, and a lower score suggests that the model tends to over-report:
    \begin{equation}
    P =
    \begin{cases}
    1, & |\mathcal{B}^{*}| = 0 \text{ and } |\mathcal{B}| = 0, \\
    0, & |\mathcal{B}^{*}| > 0 \text{ and } |\mathcal{B}| = 0, \\
    \dfrac{|\mathcal{B}_{M}|}{|\mathcal{B}|}, & |\mathcal{B}| > 0.
    \end{cases}
    \end{equation}
\end{itemize}
The per-project R-scores and P-scores are summed across the benchmark and normalized to $[0, 100]$, yielding the overall R-score and P-score.
\section{Experiments}
\paragraph{Experimental Setup} We conduct experiments with the CATJudge framework on CATTest, using the core tool set (\S\ref{sec:tools}) that includes only the Playwright script-execution tool as the GUI tool. The maximum number of interaction turns is set to $H = 250$, and both the inter-tool interval within a turn and the post-execution screenshot delay are set to 0.1\,s. For each API request, the maximum number of generated tokens is 16{,}384 and the temperature is 1.0. All the other parameters are kept at the provider's default values.

\subsection{Main Results}
Table~\ref{tab:main_results} reports the performance of mainstream VLMs (see Appendix~\ref{app:experiment_details} for the full list) on CATTest under the CATJudge framework. We can draw the following observations.

\paragraph{Even the strongest models are far from solving the CAT tasks.} The best R-score is only 42.57 (Claude-Opus-4.7) and the best P-score is only 48.56 (Gemini-3.1-Pro), meaning that no model can reliably uncover even half of the GT bugs while keeping its reports trustworthy.
Claude-Sonnet-4.6 reaches a near-perfect Recall of 98.65 yet only obtains an R-score of 37.68. GPT-5.4 exhibits the same pattern (Recall 93.67 vs.\ R-score 17.58). These models are good at flagging that ``something is wrong'' with a project, but fail to identify the specific defects that triggered the failure.

\begin{figure*}[!t]
  \centering
  \includegraphics[width=0.98\textwidth]{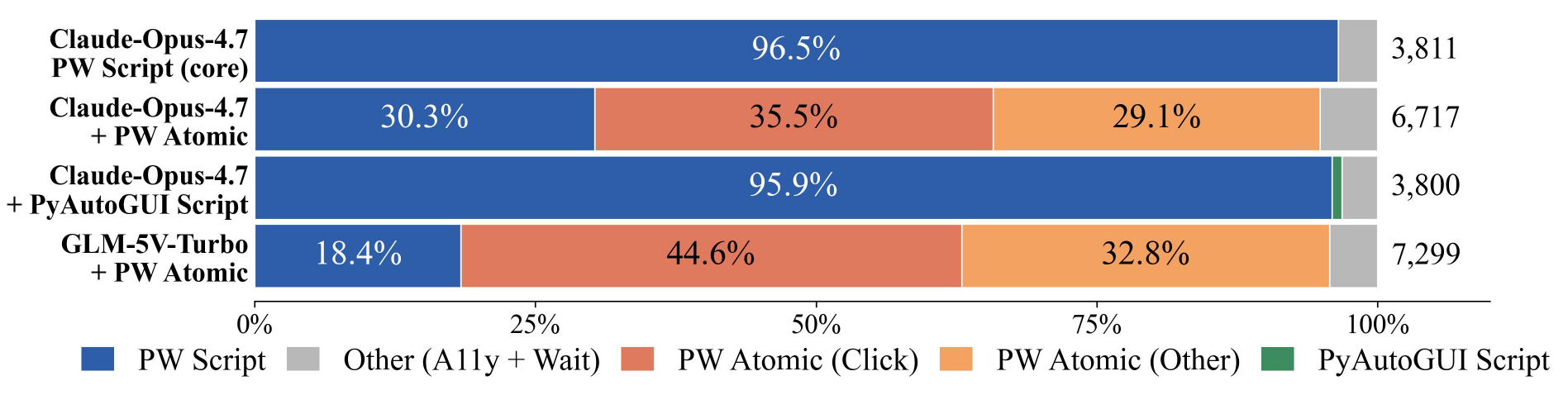}
  \caption{Tool-call distribution across the three tool-set configurations on Claude-Opus-4.7 (top three rows), together with the \emph{+ PW Atomic} configuration on the weaker GLM-5V-Turbo (bottom row) for comparison. Each horizontal bar shows the proportion of calls made to each available tool, and the number to the right of each bar reports the total number of tool calls under that configuration. \emph{PW Atomic (Click)} denotes the Playwright atomic click operation, which is reported separately because it dominates the rest of the atomic operations; \emph{PW Atomic (Other)} aggregates all remaining Playwright atomic operations.}
  \label{fig:tool_ablation}
\end{figure*}

\paragraph{A clear capability gap separates frontier models from the rest.} Five models (Claude-Opus-4.7, Claude-Opus-4.6, GPT-5.5, Claude-Sonnet-4.6, and Gemini-3.1-Pro) form a leading tier with R-score above 36. Two open models follow at an intermediate level (Qwen3.6-27B at 33.56 and Kimi-K2.6 at 30.78), after which the remaining models fall below 25, leaving a gap of roughly six points between the intermediate group and the rest. Low-capacity and lightweight variants degrade further on the coarse-grained metrics: Gemini-3.1-Flash-Lite, Doubao-Seed-2.0-Lite, and Qwen3.5-9B are the only models whose Recall falls below 41, making them unreliable for practical GUI testing. The gap is also pronounced within the same family: GPT-5.5 more than doubles GPT-5.4 in R-score (39.41 vs.\ 17.58), suggesting that the CAT task is highly sensitive to the underlying coding capability and benefits disproportionately from the latest generation of frontier models.

\subsection{Analysis}
\label{sec:analysis}
\begin{table}[t]
  \centering
  \small
  \begin{tabularx}{\columnwidth}{l *{2}{>{\centering\arraybackslash}X} @{}}
    \toprule
    \textbf{Model} & \textbf{GT} & \textbf{Reported} \\
    \midrule
    Claude-Opus-4.7  & $97.38\%$ & $96.59\%$ \\
    GLM-5V-Turbo     & $98.43\%$ & $97.10\%$ \\
    \bottomrule
  \end{tabularx}
  \caption{Agreement between Gemini-3-Flash-based automatic bug matching and human matching on the test outputs of Claude-Opus-4.7 and GLM-5V-Turbo. \textbf{GT} reports the fraction of GT bugs whose match label agrees with the human, and \textbf{Reported} reports the same on the model-reported bugs.}
  \label{tab:bug_matching}
\end{table}

\paragraph{Hybrid-Side testing matters.}
To validate our choice of Hybrid-Side testing (\S\ref{sec:catjudge}), we compare three configurations in the testing workflow on the strongest model in our pool, Claude-Opus-4.7: (1) \textit{Hybrid-Side}, the default CATJudge setup with the Playwright script-execution tool; (2) \textit{User-Side}, CATJudge with the Playwright tool replaced by PyAutoGUI; and (3) \textit{Developer-Side}, where Claude Code performs static code review without any browser interaction. As shown in Table~\ref{tab:testing_workflow}, Hybrid-Side delivers the best R-score and P-score by a clear margin. Switching to PyAutoGUI drops the R-score, since on-screen element localization becomes the bottleneck. Hybrid-Side is also cheaper than User-Side, as reported in Table~\ref{tab:efficiency} (\S\ref{app:additional_experiments}). Pure code review yields a competitive R-score but its P-score collapses, confirming that without interactive and visual feedback the model massively over-reports speculative defects that do not actually manifest at runtime.

\paragraph{Tool preference varies with model capability.}
Starting from the core tool set (\S\ref{sec:tools}), we additionally expose the Playwright atomic operations and the PyAutoGUI script-execution tool in two separate configurations, and record the resulting tool-call distributions on both Claude-Opus-4.7 and GLM-5V-Turbo (Figure~\ref{fig:tool_ablation}, full metrics in Table~\ref{tab:tool_ablation}). The strong model concentrates its calls on Playwright script-execution and seldom invokes either alternative, whereas the weaker model spreads most of its calls across the atomic operations. Beyond the relative distribution, the totals on the right of Figure~\ref{fig:tool_ablation} show that script-execution also substantially reduces the absolute number of tool calls and makes the testing more efficient.

\begin{table}[t]
  \centering
  \small
  \setlength{\tabcolsep}{4pt}
  \begin{tabular}{lcccccc}
    \toprule
    \textbf{Config} & \textbf{R} & \textbf{P} & \textbf{Acc} & \textbf{Rec.} & \textbf{Prec.} & \textbf{F1} \\
    \midrule
    Hybrid    & \textbf{42.57} & \textbf{38.30} & \textbf{83.33} & 93.67 & \textbf{86.05} & \textbf{89.70} \\
    User      & 32.94 & 27.83 & 82.35 & 94.94 & 84.27 & 89.29 \\
    Developer & 35.87 & 10.60 & 76.47 & \textbf{98.73} & 77.23 & 86.67 \\
    \bottomrule
  \end{tabular}
  \caption{Comparison of the three different testing workflow configurations evaluated on Claude-Opus-4.7. Here \textbf{R} and \textbf{P} denote the overall R-score and P-score, while \textbf{Rec.} and \textbf{Prec.} denote Recall and Precision. \textbf{Bold} marks the best result per column.}
  \label{tab:testing_workflow}
\end{table}

\section{Conclusion}
This paper studies whether VLMs can act as autonomous QA engineers for AI-generated web applications. Starting from a workflow model of agentic web generation, we address the lack of an evaluation paradigm with three contributions: a code-driven testing paradigm in which the agent treats Playwright code as its primary interface to the browser, an agentic framework that unifies BUA and CUA tools inside a single isolated environment, and a benchmark of 102 web projects with carefully annotated ground-truth bugs. Across mainstream VLMs, the highest score we measure remains well below the level required for unattended deployment, and a sizable performance cliff separates frontier models from the rest, indicating that bug discovery in real web projects is a strong probe of coding intelligence than benchmark scores on conventional code generation.

\section*{Limitations}
Our work aims to lay a foundation for code-driven agentic testing in web development. Nevertheless, several limitations remain: (1) Completeness of GT Bugs: As discussed in \S\ref{sec:benchmark_construction}, bug annotation is a highly labor-intensive process and is unlikely to be exhausted by any single effort. Despite our best efforts to identify GT bugs as comprehensively as possible, some defects inevitably remain undiscovered. Therefore, R-score does not penalize a model for reporting additional bugs on projects with bugs, and is complemented by the P-score. We expect the GT bug set to be continuously expanded through future use. (2) Evaluation of Subjective Metrics: During GT bug annotation, we clean and revise the project documents and source code to minimize subjectivity and stabilize evaluation. However, we still observe that models sometimes report minor, barely user-perceivable issues as bugs. For example, in project~99, the background animation continues to play after gameover, which is in fact common and acceptable behavior in many real-world games. In addition, to keep the evaluation as objective as possible, we focus only on functional correctness (e.g., interactive functionality, layout and rendering correctness, and performance) and do not evaluate subjective dimensions specified in web development standards, such as accessibility, user experience, and visual style. This is a deliberate scoping decision rather than a framework restriction: covering such dimensions would only require supplying additional criteria to the CATJudge agent, but it would come at the cost of evaluation stability, since models differ in how they interpret and apply subjective scoring criteria, and the resulting scores are correspondingly harder to reproduce. We therefore concentrate on functional correctness, which is the more fundamental requirement and one that current models already fail to guarantee, and we regard the evaluation of subjective dimensions as complementary work best carried out by dedicated benchmarks, so that a healthy evaluation ecosystem is built jointly from multiple efforts rather than from any single one.

\subsubsection*{Acknowledgments}
This research was supported by the National Natural Science Foundation of China (Grants No.62477044), the Key Technologies R\&D Program of Anhui Province (No.202423k09020039), the Young Elite Scientists Sponsorship Program by CAST (No. 2024QNRC001), the Fundamental Research Funds for the Central Universities (No.WK2150110038).


\bibliography{custom}

\appendix

\section{CATTest Details}
\subsection{Statistics}
\label{app:statistics}

CATTest contains $102$ AI-generated web applications, organized into $9$ top-level categories (Cat.~1--9) that are further divided into fine-grained subcategories. Each subcategory is associated with a \emph{core interaction} that characterizes its primary input modality and feedback loop. Figure~\ref{fig:category_taxonomy} summarizes the distribution of tasks across the $9$ top-level categories. To give a concrete sense of how the subcategory taxonomy is designed, Table~\ref{tab:game_subcategories} shows the full subcategory list for Cat.~8 (Game), which is the largest category in CATTest, together with the core interaction of each subcategory.

\paragraph{Prompt Complexity}
To quantify the difficulty implicitly carried by the development prompts in CATTest, we apply the same two-stage decomposition pipeline used during CATTest construction (see \S\ref{sec:benchmark_construction}) to the prompts of two open-source web generation benchmarks, WebGen-Bench~\citep{lu2026webgen} and the web subset of VIBE~\citep{vibe2025}, and count the resulting user stories per item. A larger story count indicates a richer set of latent requirements and, in turn, a more demanding generation task. We use the same backbone model, system prompts, and decoding configuration across all three datasets, so the per-item story count is directly comparable. Table~\ref{tab:prompt_complexity} reports the resulting statistics. CATTest yields the highest average number of stories per item ($43.40$), $1.43\times$ that of WebGen-Bench and $1.31\times$ that of VIBE-web. Even the prompts annotated as \emph{hard} in VIBE-web ($35.69$ stories per item) fall short of the CATTest average. Within CATTest, the most complex categories are Game (Cat.~8, $53.63$) and Web Application (Cat.~4, $52.45$), both of which involve high interaction density and rich state.

\begin{table}[t]
  \centering
  \small
  \setlength{\tabcolsep}{5pt}
  \renewcommand{\arraystretch}{1.1}
  \begin{tabularx}{0.98\columnwidth}{X r r r}
    \toprule
    \textbf{Subset} & \textbf{\#Items} & \textbf{\#Stories} & \textbf{Avg.} \\
    \midrule
    \multicolumn{4}{l}{\textit{CATTest (Ours)}} \\
    \midrule
    All                                & 102 & 4{,}427 & \textbf{43.40} \\
    \midrule
    Information Display         &  11 &   341   & 31.00 \\
    E-commerce/Transaction      &  10 &   410   & 41.00 \\
    Social Interaction          &  10 &   437   & 43.70 \\
    Web Application             &  11 &   577   & 52.45 \\
    Marketing/Landing Page      &  10 &   359   & 35.90 \\
    3D/Visualization Display    &  11 &   477   & 43.36 \\
    Data Management             &  10 &   423   & 42.30 \\
    Game                        &  19 & 1{,}019 & 53.63 \\
    SVG (DOM-Interactive)       &  10 &   384   & 38.40 \\
    \midrule
    \multicolumn{4}{l}{\textit{WebGen-Bench}} \\
    \midrule
    All                                & 101 & 3{,}072 & \textbf{30.42} \\
    \midrule
    Content Presentation               &  28 &   714   & 25.50 \\
    Data Management                    &  24 &   793   & 33.04 \\
    User Interaction                   &  49 & 1{,}565 & 31.94 \\
    \midrule
    \multicolumn{4}{l}{\textit{VIBE (web subset)}} \\
    \midrule
    All                                &  40 & 1{,}327 & \textbf{33.17} \\
    \midrule
    Easy                               &  13 &   397   & 30.54 \\
    Medium                             &  14 &   466   & 33.29 \\
    Hard                               &  13 &   464   & 35.69 \\
    \bottomrule
  \end{tabularx}
  \caption{Prompt complexity comparison across CATTest and two open-source web generation benchmarks. \textbf{\#Items} is the number of prompts; \textbf{\#Stories} is the total number of user stories obtained from the two-stage decomposition pipeline; \textbf{Avg.} is the average number of stories per item. Within each dataset, items are further grouped by their native taxonomy (top-level category for CATTest, primary category for WebGen-Bench, and difficulty level for VIBE-web).}
  \label{tab:prompt_complexity}
\end{table}

\paragraph{GT Bug Statistics}
Unlike debug-oriented benchmarks that manually inject defects into otherwise correct code, we obtain GT bugs directly from AI-coding artifacts by raising the difficulty of the generation prompts. Although we generate the web applications with Claude Code backed by Claude-Opus-4.6, which represents the frontier of AI-coding capability, and run the generation pipeline only once without iterative regeneration or self-repair, the resulting projects still expose a substantial number of bugs. Table~\ref{tab:gt_bug_statistics} summarizes the GT bug statistics. Treated as a web generation benchmark, CATTest yields a project-level pass rate (i.e., the fraction of projects whose GT bug set is empty) of only $22.55\%$ for Claude Code + Claude-Opus-4.6, with $1.86$ bugs on average per project. The bug-bearing rate is high across nearly all categories, reaching $100\%$ for Marketing/Landing Page (Cat.~5), $90.91\%$ for 3D/Visualization Display (Cat.~6), and $89.47\%$ for Game (Cat.~8). This confirms the effectiveness of our difficulty-control strategy: prompts complex enough to overwhelm a state-of-the-art coding agent naturally elicit a rich pool of real defects, removing the need for manual bug injection.

\begin{table}[t]
  \centering
  \small
  \setlength{\tabcolsep}{3pt}
  \renewcommand{\arraystretch}{1.1}
  \begin{tabularx}{0.98\columnwidth}{X c c c c}
    \toprule
    \textbf{Category} & \textbf{\#Items} & \textbf{\#B} & \textbf{B\%} & \textbf{Avg.} \\
    \midrule
    All                                & 102 & 190 & 77.45 & 1.86 \\
    \midrule
    Information Display         &  11 &   8 & 45.45 & 0.73 \\
    E-commerce/Transaction      &  10 &  20 & 70.00 & 2.00 \\
    Social Interaction          &  10 &  24 & 80.00 & 2.40 \\
    Web Application             &  11 &  32 & 72.73 & 2.91 \\
    Marketing/Landing Page      &  10 &  20 & 100.00 & 2.00 \\
    3D/Visualization Display    &  11 &  20 & 90.91 & 1.82 \\
    Data Management             &  10 &  14 & 60.00 & 1.40 \\
    Game                        &  19 &  37 & 89.47 & 1.95 \\
    SVG (DOM-Interactive)       &  10 &  15 & 80.00 & 1.50 \\
    \bottomrule
  \end{tabularx}
  \caption{GT bug statistics on CATTest, broken down by top-level category. \textbf{\#B} is the total number of GT bugs in the category; \textbf{B\%} is the fraction of projects in the category that contain at least one GT bug (i.e., $1 - $\,pass rate); \textbf{Avg.} is the average number of GT bugs per project.}
  \label{tab:gt_bug_statistics}
\end{table}

\begin{table}[t]
  \centering
  \small
  \setlength{\tabcolsep}{3pt}
  \begin{tabularx}{0.98\columnwidth}{@{\hspace{0.6em}}Xcccccc@{}}
    \toprule
    \textbf{Tool Set} & \textbf{R} & \textbf{P} & \textbf{Acc} & \textbf{Rec.} & \textbf{Prec.} & \textbf{F1} \\
    \midrule
    PW Script   & \textbf{42.57} & 38.30          & \textbf{83.33} & \textbf{93.67} & 86.05          & \textbf{89.70} \\
    + PW Atomic        & 35.07          & 30.20          & 82.35          & 89.87          & \textbf{87.65} & 88.75 \\
    + PA Script & 40.13          & \textbf{43.24} & 81.37          & 91.14          & 85.71          & 88.34 \\
    \bottomrule
  \end{tabularx}
  \caption{Tool set ablation evaluated on Claude-Opus-4.7. \emph{PW Script (core)} is the core tool set used in the main experiments. \emph{+ PW Atomic} additionally exposes the Playwright atomic operations; \emph{+ PA Script} additionally exposes the PyAutoGUI script-execution tool. \textbf{R} and \textbf{P} denote the overall R-score and P-score, while \textbf{Rec.} and \textbf{Prec.} denote Recall and Precision. \textbf{Bold} marks the best result per column.}
  \label{tab:tool_ablation}
\end{table}

\subsection{Models Involved in Bug Annotation}
\label{app:model_in_annotation}
During the bug annotation stage of the benchmark construction, we adopt a tool-competition pooling strategy. To this end, we run several strong VLMs on CATJudge to discover bugs. The models involved in this process are:
Claude-Opus-4.7, Claude-Opus-4.6, Gemini-3.1-Pro, Gemini-3-Flash, Gemini-3.1-Flash-Lite, GPT-5.5, GLM-5V-Turbo, and Qwen3.5-Plus.

\subsection{On the Choice of Code Generator}
\label{app:generator_choice}
The evaluation scope of CATTest is deliberately restricted to \emph{subtle, naturally occurring} bugs in coding-agent scenarios (\S\ref{sec:benchmark_construction}), and our use of a frontier coding agent as the generator is a design decision aligned with that scope.

The purpose of the generation stage is to raise the difficulty of the \emph{debugging} task. A weaker generator would work against this purpose in two ways. First, it lowers the ceiling of bug subtlety: the defects it leaves behind tend to be coarse and immediately visible, so every tester scores high and the benchmark loses its power to discriminate among testers of different strength. Second, and more damaging, weaker generators frequently fail before the interactive layer is ever reached. In our preliminary trials with less capable generator configurations, a large fraction of the resulting projects did not compile, crashed on start-up, or failed to launch at all because of syntax and configuration errors. Such projects are trivially rejected by any tester without a single meaningful interaction, and they are therefore useless for evaluating interaction-driven bug discovery, which is precisely the ability CATTest is designed to measure. Using a frontier generator instead yields projects that run correctly, look plausible, and fail only under specific interaction sequences, which is the regime of interest.

Two further points bound the influence of this choice on the validity of the evaluation. First, the evaluation is anchored on the GT bug set rather than on the generator: the generation prompts are never exposed to the models under test, which only observe a fixed, already-running web project, and the GT bug set is produced by a tool-competition pooling strategy that combines human verification with several strong models from different families (\S\ref{app:model_in_annotation}). Second, a tester that misses a GT bug reveals a genuine capability deficiency, independently of whether the same defect would also arise in code produced by some other generator. Finally, the construction pipeline (Figure~\ref{fig:benchmark_construction}) is fully specified and extensible: alternative generators can be substituted and additional projects annotated under the same protocol, so CATTest can be extended along whichever axis a future evaluation goal requires.

\subsection{Bug Matching Criterion}
\label{app:bug_matching}
Both the annotation stage (\S\ref{sec:benchmark_construction}) and the fine-grained metrics (\S\ref{eval}) rely on deciding whether a reported bug and a GT bug refer to the same underlying defect. Because a single defect can surface as several distinct user-visible symptoms, matching cannot be reduced to textual similarity. We therefore adopt the following criterion. A pair of bugs is considered a match if and only if at least one of the following conditions holds:
\begin{enumerate}[leftmargin=*,itemsep=2pt,topsep=2pt]
  \item the two descriptions refer to essentially the same functional problem, even when worded differently;
  \item the first describes a symptom or behavior that is directly caused by the same underlying defect as the second;
  \item the first describes a consequence of the second, i.e.\ it would not occur if the second were fixed.
\end{enumerate}
Matching is performed as an all-pairs comparison between the reported bug list and the GT bug list, with the criterion applied to each pair independently; each reported bug is accompanied by a description, reproduction steps, and expected versus actual behavior (Figure~\ref{fig:prompt_qa_explorer}), which supplies the evidence needed to decide the relation. The resulting matched-pair set is the set $M$ used in \S\ref{eval}.

\paragraph{Partial matches and supersets.} The two lists need not describe a defect at the same granularity, and the resulting containment can run in either direction: a single reported bug may cover several GT bugs, or a single GT bug may be split across several reported bugs. Either way, condition~(1) is applied part-wise, so a bug that captures only a component of its counterpart still matches it. Project~38, a cloud-server terminal emulator, illustrates the first direction. Two GT bugs, \emph{``cursor position deviates from the command-line prompt''} and \emph{``after input, the text position shifts upward relative to the cursor position''}, are both judged to match the single reported bug \emph{``input line broken: typed text and cursor render far away from the prompt and on separate lines''}, since each GT bug describes one component of the broken layout that the report states as a whole. The converse case, in which one GT bug is only jointly covered by several reported bugs, is handled symmetrically.

\paragraph{Causal relationships.} The two lists may also describe a defect at different points along the same causal chain, and again either side may be the deeper one. GT bugs are first recorded from the user's perspective during the human experience step (\S\ref{sec:benchmark_construction}), whereas a tester that inspects the page source is free to report the underlying root cause directly; conversely, a tester may report only a downstream symptom of a GT bug stated at its root. Conditions~(2) and~(3) cover these two directions symmetrically, so a match does not require the two descriptions to sit at the same depth. Project~58, a 3D building-model visualization, illustrates the case where the report is the deeper of the two: the GT bug \emph{``measurement numbers can only be cleared using the Clear All button''} is matched by the reported bug \emph{``per-measurement delete ($\times$) button is unreachable due to \texttt{pointer-events: none} on the label''}, because the GT symptom is a direct consequence of the reported root cause. Had the report instead noted only that individual measurements cannot be removed, it would match the same GT bug in the opposite direction.

\paragraph{A counterexample.} Neither direction may be granted on superficial plausibility alone: the causal link has to actually hold. In the \emph{Insufficient Interaction Coverage} case (\S\ref{app:error_patterns}), the reported \emph{``Delete Node does not work''} reads as a downstream symptom of the GT bug \emph{``single-click cannot select a node''}, since a node that cannot be selected cannot be deleted, which would make it a match under condition~(3). The deletion functionality is nevertheless correct: a node can still be selected by a box-select drag and then deleted successfully. The apparent dependency is thus an artifact of the report being written from an incomplete exploration rather than a real causal relation, so the two bugs are \emph{not} matched, and the GT bug counts as missed.

\begin{figure*}[t]
  \centering
  \begin{minipage}{0.48\textwidth}
    \centering
    \includegraphics[width=\linewidth]{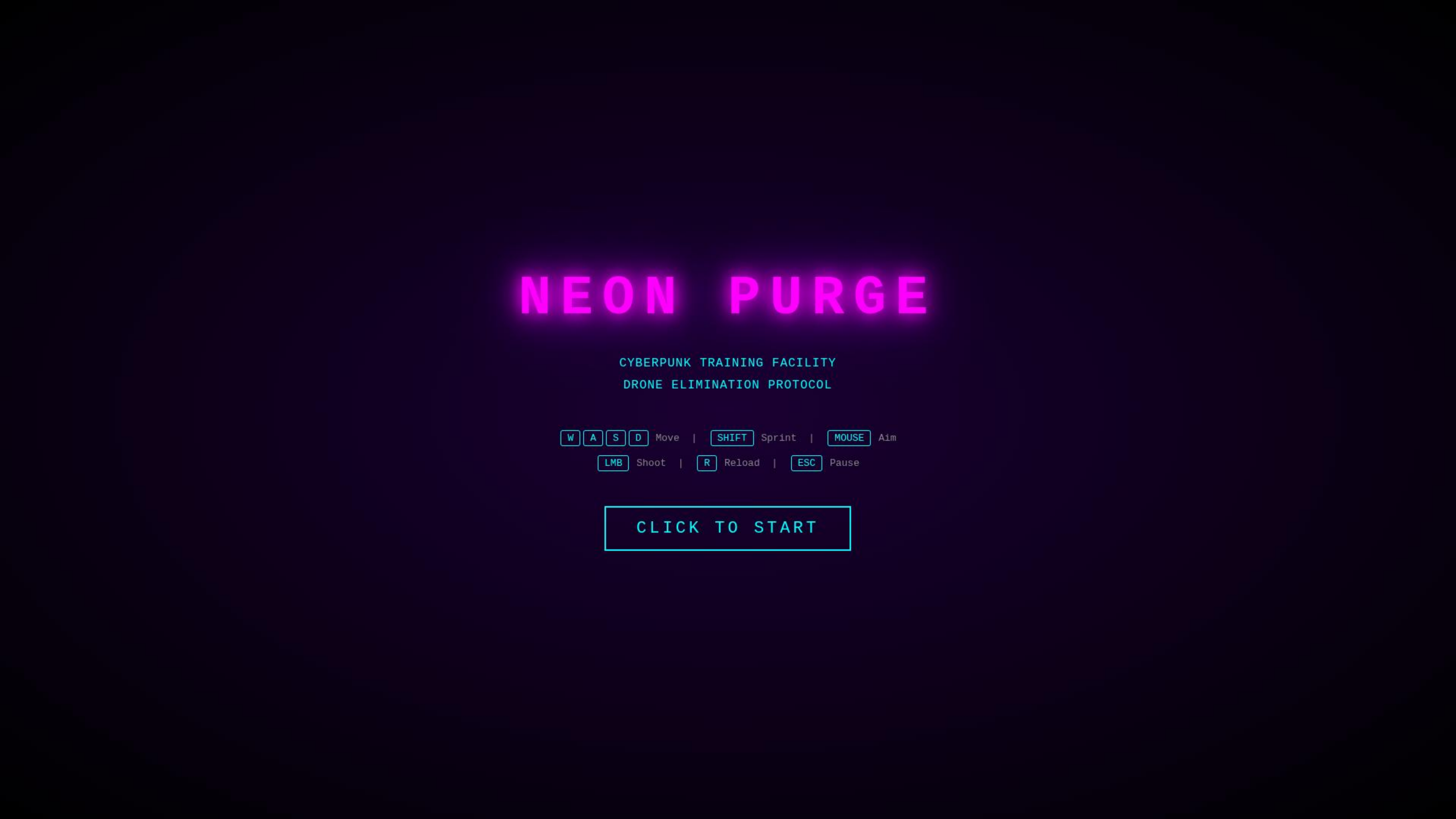}
    \subcaption{Start screen with the title "NEON PURGE".}
    \label{fig:example_screenshots_initial}
  \end{minipage}
  \hfill
  \begin{minipage}{0.48\textwidth}
    \centering
    \includegraphics[width=\linewidth]{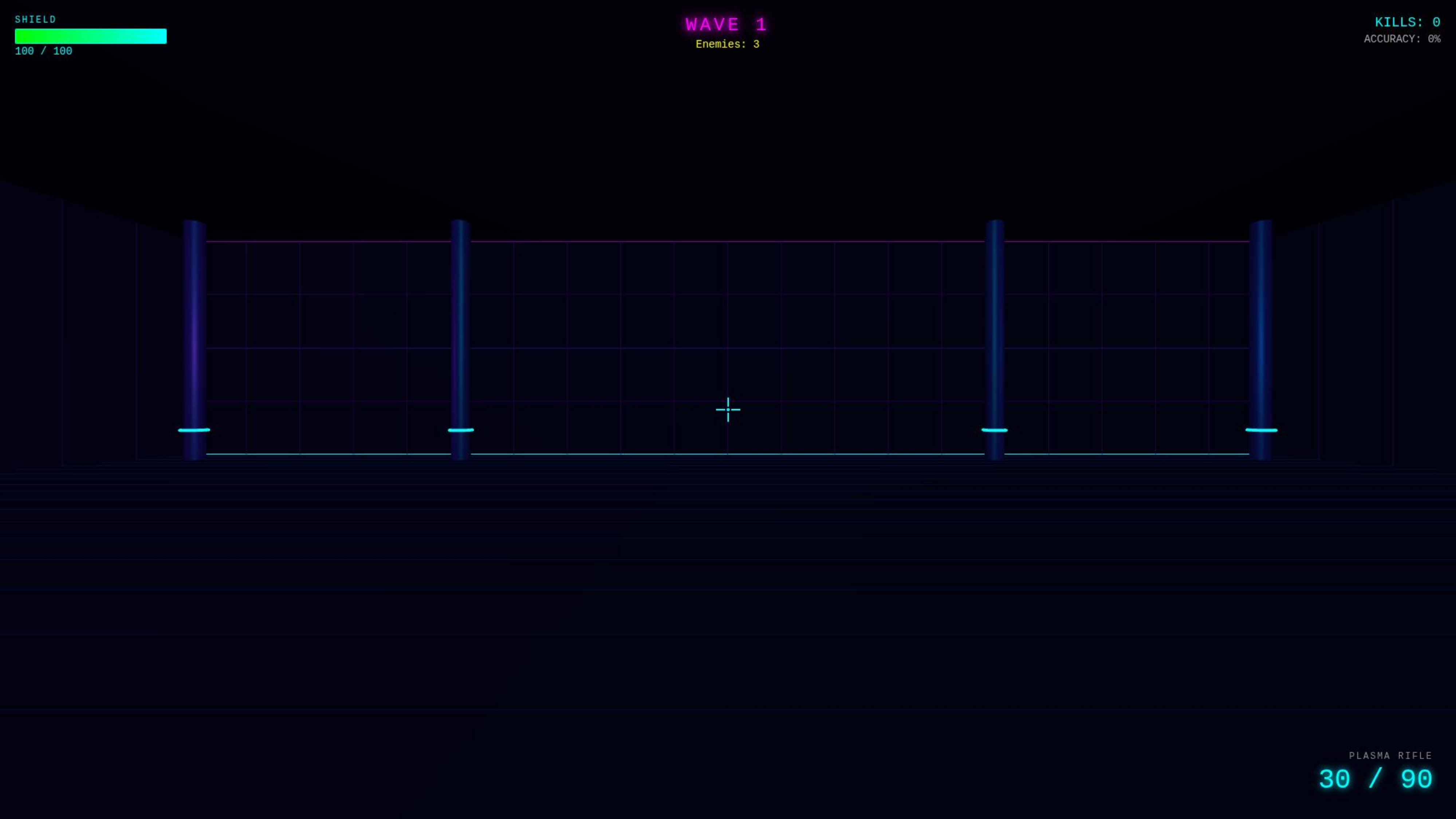}
    \subcaption{In-game state during a wave, showing the cyberpunk arena.}
    \label{fig:example_screenshots_gameplay}
  \end{minipage}
  \caption{Screenshots of the web application generated for Project~71 (\emph{Neon Purge}, Cat.~8 / FPS-Shooter).}
  \label{fig:example_screenshots}
\end{figure*}

\subsection{Example}
\label{app:example}

To give a concrete sense of what a CATTest data point looks like, we use Project~71 (\emph{Neon Purge}) as a running example. This project belongs to top-level category Cat.~8 (Game) and the subcategory \emph{FPS / Shooter}, whose core interaction is \emph{Mouse aim + left-click shoot, WASD move, crosshair feedback, ammo management} (See Table~\ref{tab:game_subcategories}). The example consists of four parts: (i) the development prompt sent to Claude Code (Figure~\ref{fig:example_prompt}); (ii) the QA-oriented project document handed off to the agent under test (Figure~\ref{fig:example_projdoc}); (iii) screenshots of the resulting web application (Figure~\ref{fig:example_screenshots}); and (iv) the GT bugs list.

\paragraph{GT Bugs.} Project~71 contains a single GT bug:
\begin{quote}
\textit{Pausing during the countdown between waves does not interrupt the countdown, and the pause overlay gets stuck on screen and does not disappear.}
\end{quote}
This defect manifests only at the boundary between two game states (\textsc{wave\_transition} $\rightarrow$ \textsc{countdown} $\rightarrow$ \textsc{playing}). Reproducing it requires the agent to actively drive the game into the wave-transition window and press \texttt{Esc} at the right moment, illustrating the kind of subtle, interaction-dependent defects that motivate CAT.

\section{Experiment Details}
\label{app:experiment_details}
\paragraph{VLMs involved in the experiments} We aim to cover the strongest model released by each major provider. To be eligible for our evaluation, a model must possess image-understanding capability. The full list of models is as follows:
\begin{itemize}
    \item Claude series: Claude-Opus-4.7~\citep{anthropic2026claudeopus47}, Claude-Opus-4.6~\citep{anthropic2026claudeopus46}, Claude-Sonnet-4.6~\citep{anthropic2026claudesonnet46}.
    \item Gemini series: Gemini-3.1-Pro~\citep{google2026geminipro}, Gemini-3-Flash~\citep{google2025gemini3flash}, Gemini-3.1-Flash-Lite~\citep{google2026gemini31flashlite}.
\item GPT series: GPT-5.5~\citep{openai2026gpt55}, GPT-5.4~\citep{openai2026gpt54}, GPT-5.4-mini~\citep{openai2026gpt54mininano}.
    \item Qwen series: Qwen3.5-Plus~\citep{qwen2026qwen35}, Qwen3.6-27B~\citep{qwen3.6-27b}, Qwen3.5-27B~\citep{qwen3.5}, Qwen3.5-9B~\citep{qwen3.5}.
    \item Others: GLM-5V-Turbo~\citep{hong2026glm}, Doubao-Seed-2.0-Pro~\citep{bytedance2026doubaoseed2}, Doubao-Seed-2.0-Lite~\citep{bytedance2026doubaoseed2}, Kimi-K2.6~\citep{moonshot2026kimik26}.
\end{itemize}

\section{Additional Experiments}
\label{app:additional_experiments}

\paragraph{Evaluation is Stable on CATTest.}
To assess the run-to-run stability of CATTest under our default setting, we repeat the full evaluation of Claude-Opus-4.7 five times and report the mean and standard deviation of every metric in the \emph{w/ doc} rows of Table~\ref{tab:doc_ablation}. These numbers indicate that CATTest yields reproducible measurements across independent runs of the same model.

\paragraph{Basic documental guidance is important.}
We ablate the project documentation on Claude-Opus-4.7. Comparing the two rows of Table~\ref{tab:doc_ablation}, the coarse-grained metrics move only modestly. The fine-grained R-score and P-score, however, fall sharply from $40.64$ to $27.17$ and from $38.79$ to $23.25$, a gap of more than five times the corresponding standard deviations. This pattern supports our design choice of providing minimal documental guidance.

\begin{table}[t]
  \centering
  \small
  \setlength{\tabcolsep}{4pt}
  \begin{tabular*}{0.98\columnwidth}{@{\hspace{0.8em}}l@{\extracolsep{\fill}}ccc@{}}
    \toprule
    \textbf{Setting} & \textbf{R} & \textbf{P} & \textbf{Acc} \\
    \midrule
    w/ doc  & $40.64_{\scriptscriptstyle\pm 2.19}$ & $38.79_{\scriptscriptstyle\pm 2.73}$ & $80.78_{\scriptscriptstyle\pm 1.91}$ \\
    w/o doc & $27.17$ & $23.25$ & $78.43$ \\
    \midrule
                    & \textbf{Rec.} & \textbf{Prec.} & \textbf{F1} \\
    \midrule
    w/ doc  & $89.87_{\scriptscriptstyle\pm 4.29}$ & $86.02_{\scriptscriptstyle\pm 1.43}$ & $87.84_{\scriptscriptstyle\pm 1.52}$ \\
    w/o doc & $93.67$ & $81.32$ & $87.06$ \\
    \bottomrule
  \end{tabular*}
  \caption{Evaluation on Claude-Opus-4.7. The \emph{w/ doc} rows report mean${}_{\pm\text{std}}$ over five runs of the default setting; the \emph{w/o doc} rows report a single run with the in-context project documentation removed. \textbf{R} and \textbf{P} denote the overall R-score and P-score; \textbf{Rec.} and \textbf{Prec.} denote Recall and Precision.}
  \label{tab:doc_ablation}
\end{table}

\paragraph{Hybrid-Side testing is also cheaper.}
Beyond the accuracy gap reported in Table~\ref{tab:testing_workflow}, we compare the Hybrid-Side and User-Side configurations in terms of testing cost on Claude-Opus-4.7. Table~\ref{tab:efficiency} reports the average wall-clock time to complete the testing of one project, together with the total number of API calls, the total number of tokens consumed, and the total inference cost aggregated over the whole benchmark. Hybrid-Side testing is not only more accurate but also uniformly cheaper along every axis: it finishes a project in $9.1$ minutes on average against $13.9$ minutes for User-Side, and consumes roughly $65\%$ of the API calls, $45\%$ of the tokens, and $58\%$ of the monetary cost. The gap follows directly from the interaction granularity of the two paradigms. A single Playwright script can express a compound interaction and return an aggregated observation in one turn, whereas driving the same interaction through keyboard and mouse primitives requires many turns, each paying the cost of a full model call and each carrying an additional screenshot into the context. Code as an interface is thus the more economical modality as well as the more capable one.

\begin{table}[t]
  \centering
  \small
  \setlength{\tabcolsep}{4pt}
  \begin{tabularx}{0.98\columnwidth}{X *{4}{>{\centering\arraybackslash}X}}
    \toprule
    \textbf{Config} & \textbf{Time} & \textbf{Calls} & \textbf{Tokens} & \textbf{Cost} \\
    \midrule
    Hybrid & \textbf{9.1}  & \textbf{3{,}912} & \textbf{0.89} & \textbf{181} \\
    User   & 13.9          & 5{,}982          & 1.96          & 311 \\
    \bottomrule
  \end{tabularx}
  \caption{Testing cost of the Hybrid-Side and User-Side configurations on Claude-Opus-4.7. \textbf{Time} is the average wall-clock time spent on one project, in minutes; \textbf{Calls} is the total number of API calls over the benchmark; \textbf{Tokens} is the total number of tokens consumed, in billions; \textbf{Cost} is the total inference cost in USD. \textbf{Bold} marks the better result per column.}
  \label{tab:efficiency}
\end{table}

\section{Case Study}
\subsection{Good Cases}
\label{app:case_good}

A useful way to see why a code-driven testing paradigm matters is to consider the projects that are \textbf{nearly impossible} to test through atomic mouse and keyboard Playwright operations alone. Browser games rendered on a single \texttt{canvas} surface, or on WebGL, are the clearest example: from the outside, the page exposes essentially one large opaque rectangle, with no individually addressable buttons, lists, or text fields for an atomic testing tool to target (for example, a single \texttt{<canvas>} node in the a11y tree, without any semantic information). A bug that only appears at the two-minute mark of a shoot-em-up, or in a transition between game modes, demands that the tester actually \emph{play} the game to that point, and that often means surviving randomized enemy waves, collecting the right power-ups, and reaching specific boss states. For an AI agent, this is both slow and unreliable: a single attempt at a late-game state can take many minutes of real-time play and depends on stochastic events the agent has no stable way to control with GUI actions alone. Sometimes a human can have trouble playing a game. The natural alternative is to have the agent write code that interacts with the application's internal logic, rather than chase the moving target through the rendered surface.

\begin{figure*}[t]
  \centering
  \includegraphics[width=0.98\textwidth]{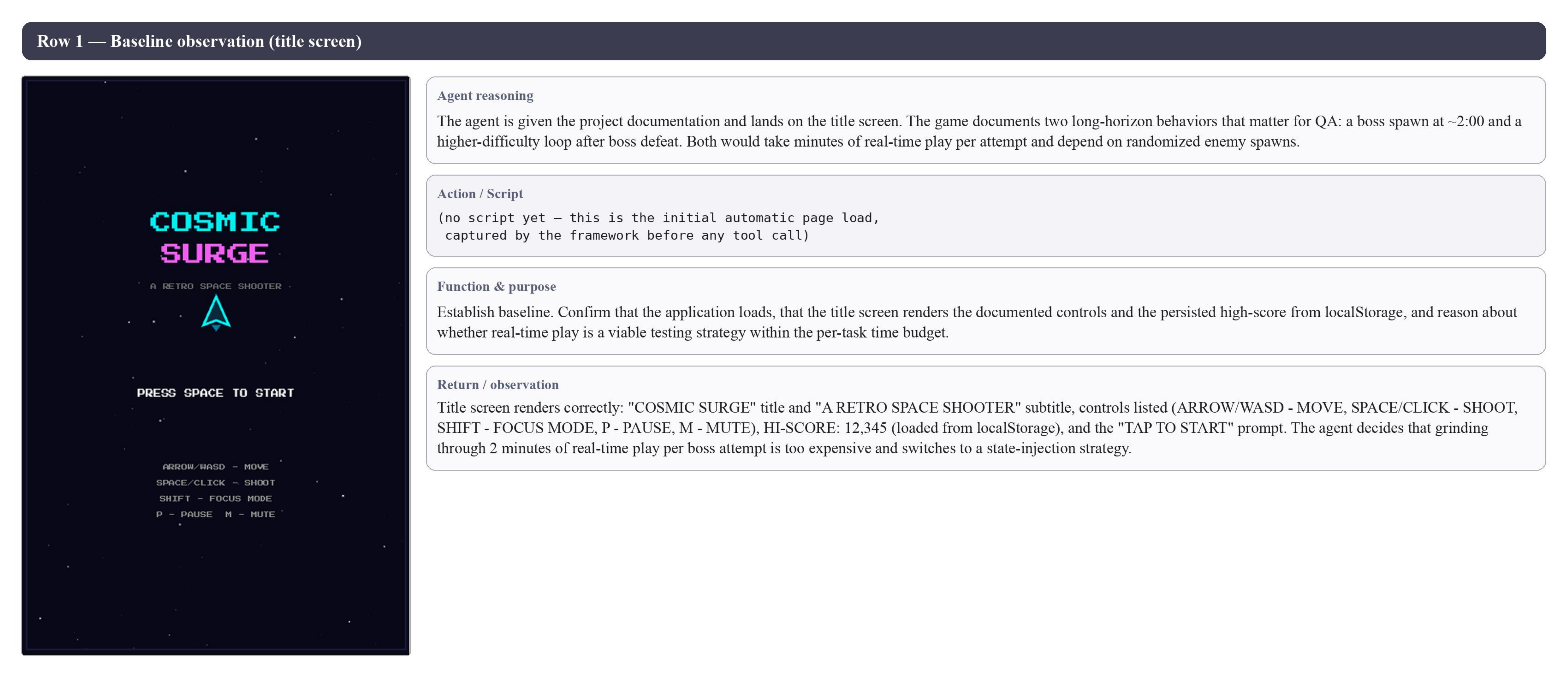}
  \caption{Project~99 (\emph{Cosmic Surge}), turn~1: baseline observation. The agent reads the project documentation and lands on the title screen, recognises that the documented long-horizon behaviours (boss spawn at 2:00, harder loop after boss defeat) are essentially unreachable through real-time play under the per-task time budget, and decides to bypass real-time play.}
  \label{fig:case_cs99_row1}
\end{figure*}

\begin{figure*}[t]
  \centering
  \includegraphics[width=0.98\textwidth]{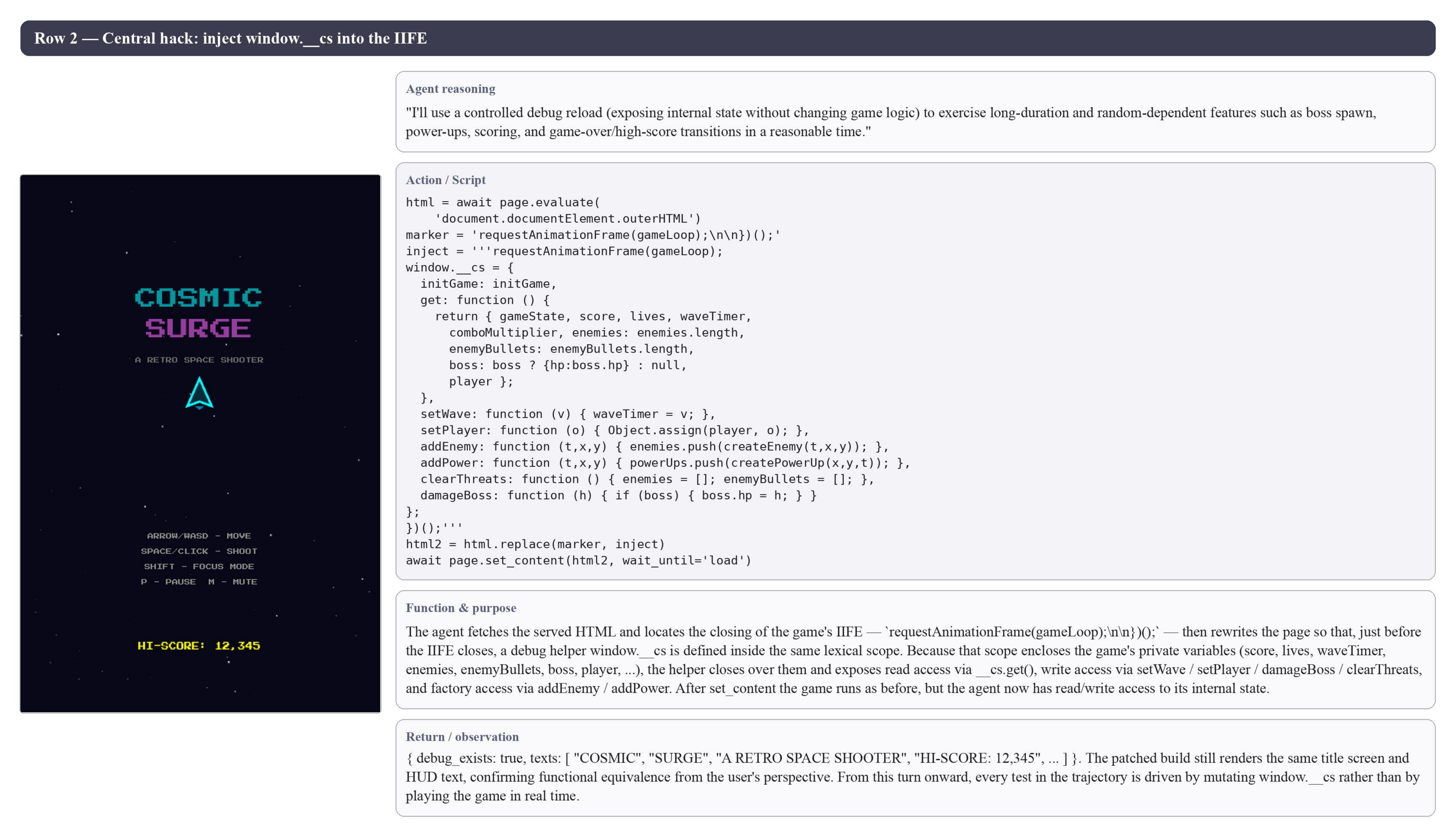}
  \caption{Project~99 (\emph{Cosmic Surge}), turn~2: the central hack. The agent fetches the served HTML, locates the closing of the game's IIFE (\texttt{requestAnimationFrame(gameLoop);\textbackslash n\textbackslash n\})();}), and rewrites the page so that just before the IIFE closes a debug helper \texttt{window.\_\_cs} is defined inside the same lexical scope. Because that scope encloses the game's private state, the helper closes over it and exposes both read access (\texttt{get()}) and write access (\texttt{setWave}, \texttt{setPlayer}, \texttt{damageBoss}, \texttt{clearThreats}, \texttt{addEnemy}, \texttt{addPower}). After \texttt{set\_content}, every subsequent test in the trajectory is driven by mutating \texttt{window.\_\_cs} rather than by playing the game in real time.}
  \label{fig:case_cs99_row2}
\end{figure*}

\begin{figure*}[t]
  \centering
  \includegraphics[width=0.98\textwidth]{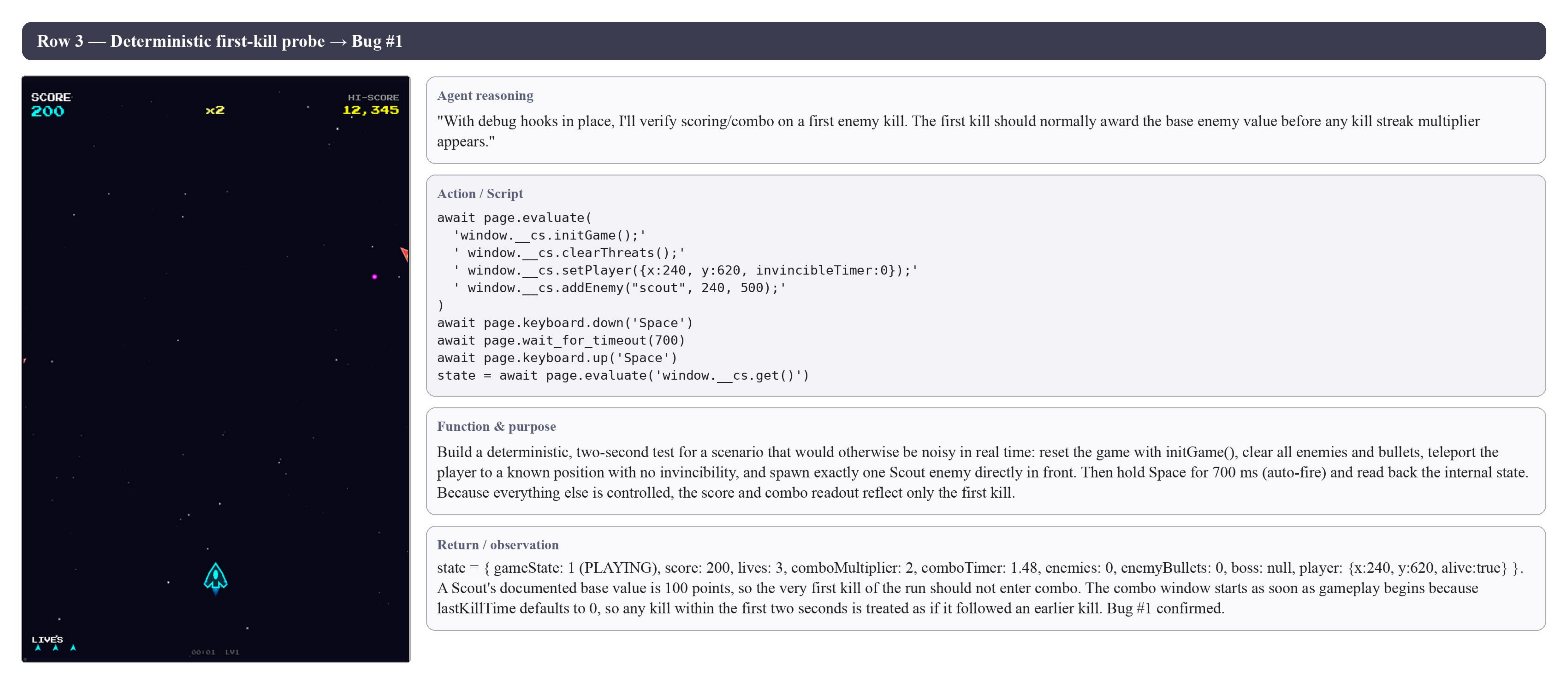}
  \caption{Project~99 (\emph{Cosmic Surge}), turn~3: a deterministic first-kill probe driven through the helper. The agent resets the game, clears all threats, teleports the player to a known position, and spawns exactly one Scout. After a 700\,ms auto-fire burst, \texttt{score=200} and \texttt{comboMultiplier=2} on the very first kill, but a Scout's documented base value is~100 and combo should not start before any prior kill, exposing a scoring bug whose root cause is \texttt{lastKillTime=0} at \texttt{initGame()}. (Bug~\#1.)}
  \label{fig:case_cs99_row3}
\end{figure*}

\begin{figure*}[t]
  \centering
  \includegraphics[width=0.98\textwidth]{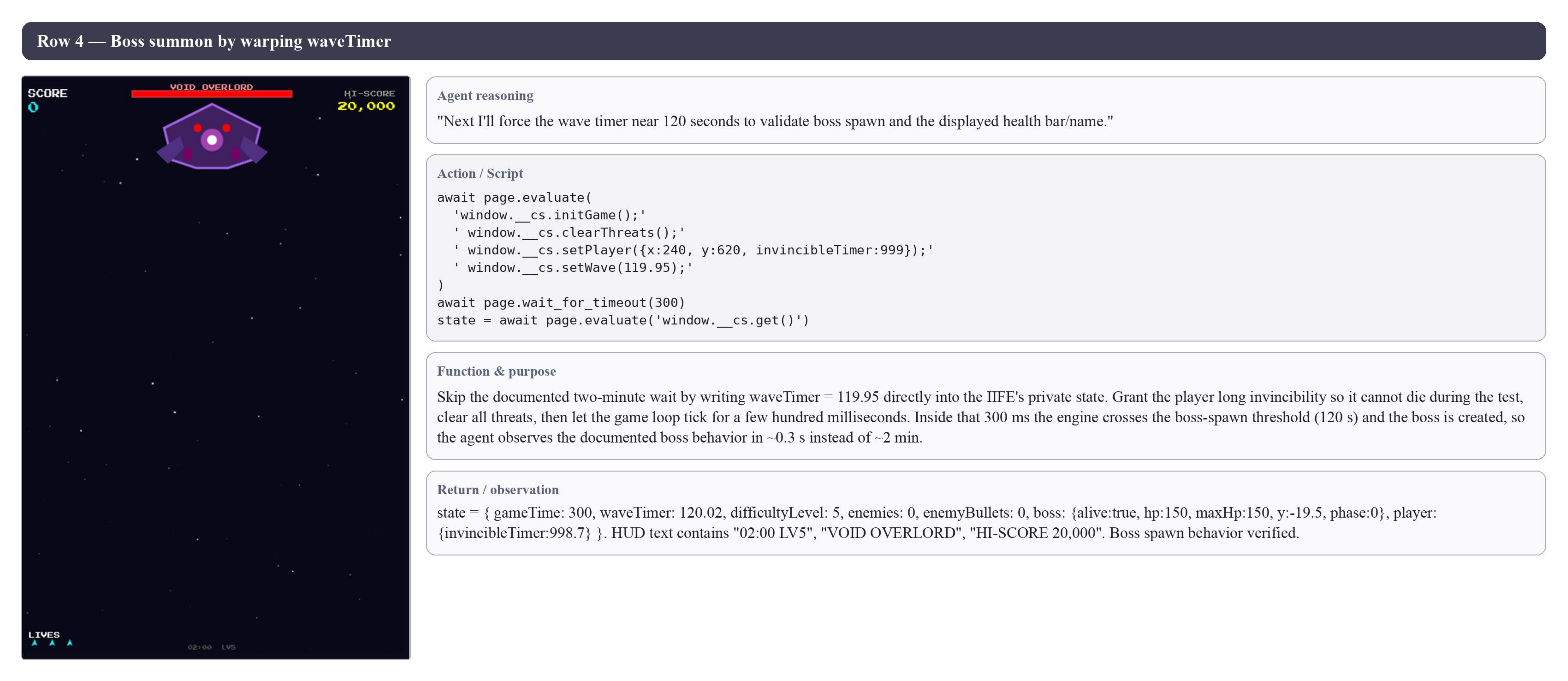}
  \caption{Project~99 (\emph{Cosmic Surge}), turn~4: skipping the documented two-minute wait by writing \texttt{waveTimer = 119.95} directly. After the game loop ticks for 300\,ms, the boss-spawn condition fires, and the HUD renders \texttt{02:00 LV5} and \texttt{VOID OVERLORD} with full HP: the documented boss behaviour is now observable in 0.3\,s instead of 2\,min.}
  \label{fig:case_cs99_row4}
\end{figure*}

\begin{figure*}[t]
  \centering
  \includegraphics[width=0.98\textwidth]{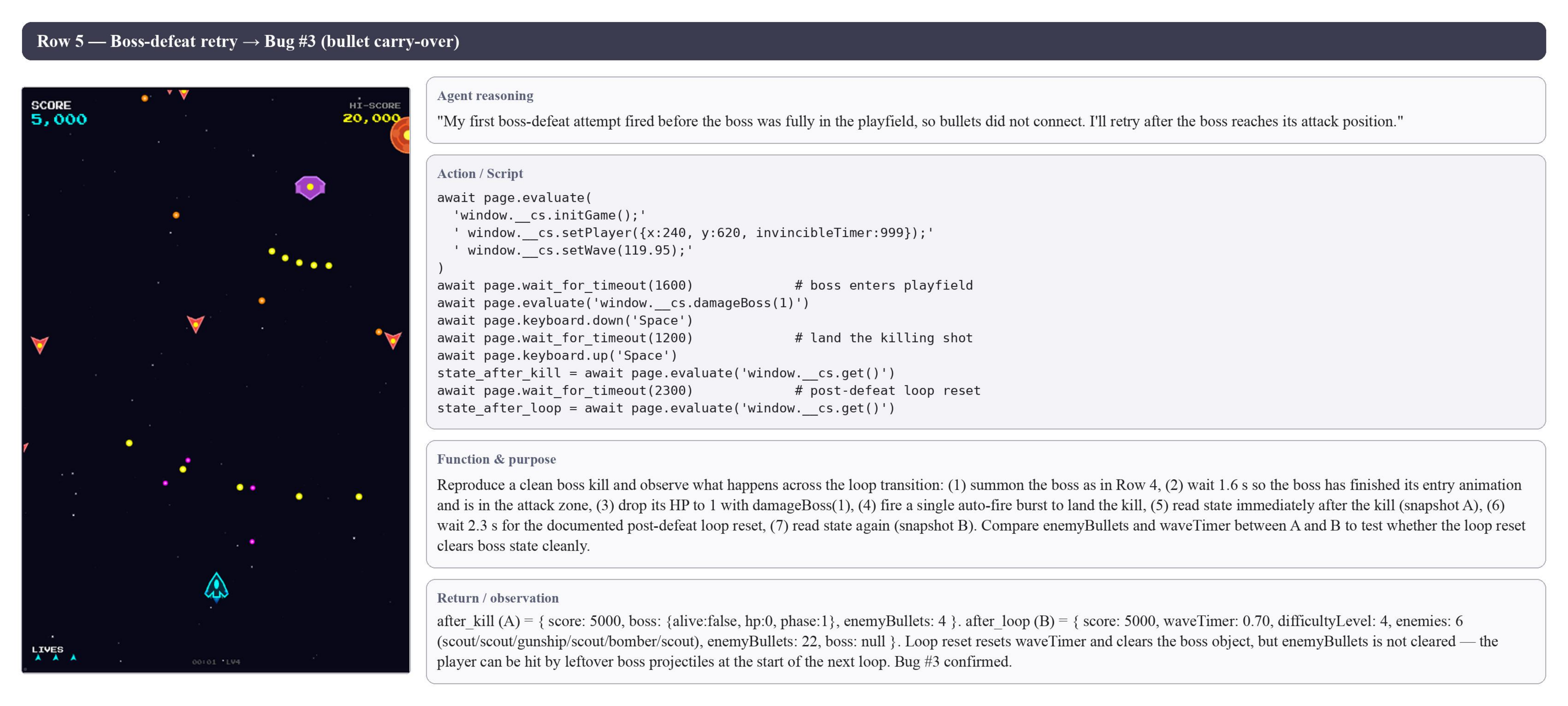}
  \caption{Project~99 (\emph{Cosmic Surge}), turn~5: a clean boss kill driven through the helper, followed by a comparison across the post-defeat loop reset. Snapshot~A (immediately after the kill) shows \texttt{score=5000, boss.alive=false, enemyBullets=4}; snapshot~B (after the documented 2.3\,s loop transition) shows \texttt{waveTimer=0.7, difficultyLevel=4} and a fresh wave of enemies, but \texttt{enemyBullets=22}: the loop reset clears the boss but leaves its projectiles in flight, so the next loop starts with leftover bullets. (Bug~\#3.)}
  \label{fig:case_cs99_row5}
\end{figure*}

The good case below illustrates exactly this. We walk through a representative trajectory produced by GPT-5.5 on Project~99 (\emph{Cosmic Surge}, a 2D shoot-em-up). Two of the project's documented behaviors, a boss that spawns around the two-minute mark and a higher-difficulty loop that begins after the boss is defeated, are practically out of reach for play-only exploration. Instead of playing, the agent treats the page itself as a programmable artifact. Powered with free-access to Playwright script-execution, it first downloads the page's source code, then writes a small piece of JavaScript that installs a backdoor into the game's internal state and re-loads the page with that backdoor in place. Once the backdoor is installed, the agent no longer has to play: it can directly read out variables that are normally private (the current wave timer, the boss HP bar, the active enemies on screen) and overwrite them at will (jumping the timer forward, spawning enemies at chosen positions, draining boss HP to one shot). All bug detections in this run, namely an inflated first-kill combo counter and a boss bullet that incorrectly carries over into the next loop, are obtained through this approach rather than by normally playing the game. We walk through the key turns of the trajectory in Figures~\ref{fig:case_cs99_row1}--\ref{fig:case_cs99_row5}. For each turn we show, side-by-side, the screenshot captured after the action, the agent's verbatim reasoning, the script it executed, a natural-language summary of the script's purpose, and the structured return values it observed.

\subsection{Error Patterns}
\label{app:error_patterns}

\paragraph{Mirage Reasoning.} A recurring failure mode is \emph{mirage reasoning} (e.g., the gent tends to ignore visual information or ignore textual information~\citep{asadi2026mirage}): the agent talks itself into a coherent story about why a strange observation is fine, and then treats that story as ground truth, instead of returning to the visual evidence. Figure~\ref{fig:error_mirage_reasoning} illustrates the pattern on Project~45, a flipping-card countdown timer. The screenshot shows the digits visibly malformed: each cell, which is meant to hold a single digit, is showing two digits stacked on top of each other. After clearing routine checks elsewhere on the page, the agent fast-forwards the timer and reads back the contents of the digit cells, finding roughly twice as many digits as the design calls for. Rather than weigh that count against the static screenshot in front of it, the agent decides the extra digits must be a fleeting frame of an in-progress flip animation; a follow-up check that merely confirms an animation is happening is then taken as proof that the timer is fine. The verdict is \textsc{pass} with zero bugs. The failure is not perceptual, the broken digits are plainly visible, but a willingness to prefer a fluent explanation (``a mid-flip frame'') to the simpler reading that the page is showing twice as many digits as it should.

\paragraph{Insufficient Interaction Coverage.} A second pattern is what we call \emph{insufficient interaction coverage}: the agent reaches for high-level shortcuts such as ``click this'' or ``press that'' by default, and never constructs the more elaborate gestures that the project's edge-case bugs actually require, or would help to reveal the true bug. Figure~\ref{fig:error_interaction_coverage} illustrates the pattern on Project~34 (\emph{FlowWeave}, a node editor). Three of the four documented bugs are gesture-shaped: dragging a node and releasing the mouse over the right-hand panel makes the node keep following the cursor; with snap enabled the drag visibly lags behind the cursor; and a single click cannot select a node, only the box-select drag does. Across the run the agent issues a long string of high-level interactions but only a handful of true mouse-press-move-release gestures, all of them safely contained inside the canvas; it never tries finishing a drag outside the canvas, never re-runs a drag with snap turned on, and never compares plain clicking against the alternative selection paths. As a result the report flags five surface anomalies, but only one of them lines up with a real bug (the right-click ``Delete Connection'' being inert). The agent even writes that ``no node has visual selection highlight,'' a single follow-up step away from discovering that single-click does not select at all, which is the true bug leading to the observed "Delete Node does not work" phenomenon (the ``Delete Node'' function itself works as expected).

\paragraph{Spec Hallucination.} A third pattern, complementary to mirage reasoning, is what we call \emph{spec hallucination}: the agent imports its prior beliefs about how a class of applications ``ought to'' behave and treats those beliefs as binding requirements, even when the project's documentation says nothing of the kind, and even when the agent's own follow-up observations contradict the invented rule. Figure~\ref{fig:error_spec_hallucination} illustrates the pattern on Project~13 (\emph{BloomBox}, a gardening e-commerce site), a project that carries no documented bugs at all. After completing a normal checkout, the agent notices that the cart counter still reads ``7'' on the order-confirmation page and writes ``the cart should be cleared after a successful order'', a rule the project documentation does not state. In the very next turn the agent observes that the cart \emph{is} cleared after one further click, and yet keeps the complaint in its bug list, downgrading it to ``clearing happens late.'' A second invented rule, that stock must decrement after a sale, survives even the agent's own aside that the unchanged stock count ``may be by design.'' A third report flags ellipsis-shortened product names as a cosmetic defect even though that layout is clearly intentional. The result is three false-positive bugs and a \textsc{not\_pass} verdict on a clean project. The failure is not weak observation; it is a category mismatch between the agent's defaults for generic e-commerce sites and the project's actual contract.

\section{Prompts}

This section collects the prompts referenced throughout the paper, reproduced verbatim so that all design choices are inspectable.

\section{Human Annotators}
The annotators are MS and PhD students with a general CS background, with one author serving as the team leader responsible for organizing the annotation process. Annotation tasks are evenly distributed among the annotators, and the annotations are used only after being approved by the team leader. Each annotator is paid a reasonable salary, and all costs of AI tools used during annotation are borne by the author team. AI tools are employed not only to assist the annotators but also as static scripts to independently verify the data (see \S~\ref{sec:cattest}).

\section{Use of LLMs}
We use publicly available LLMs to assist and polish our writing.

\begin{table*}[t]
  \centering
  \small
  \setlength{\tabcolsep}{4pt}
  \renewcommand{\arraystretch}{1.15}
  \begin{tabular}{p{0.30\textwidth} p{0.62\textwidth}}
    \toprule
    \textbf{Subcategory} & \textbf{Core Interaction} \\
    \midrule
    FPS / Shooter & Mouse aim + left-click shoot, WASD move, crosshair feedback, ammo management. \\
    \midrule
    Board / Turn-Based Strategy & Click piece $\rightarrow$ click target cell, turn waiting, legal move highlighting. \\
    \midrule
    Platformer & Arrow-key move + space jump, gravity physics, platform collision. \\
    \midrule
    Match / Elimination Puzzle & Click or drag to swap adjacent elements, chain elimination animation, scoring. \\
    \midrule
    Maze / Stealth & Keyboard directional movement, fog-of-war/limited-visibility exploration, avoid patrol paths, stealth-state detection. \\
    \midrule
    Tower Defense / Placement & Drag-place defense units on grid, upgrade click, wave progression. \\
    \midrule
    Text Adventure / Choice & Read text $\rightarrow$ click choice options, branching storyline, save management. \\
    \midrule
    Physics Puzzle & Drag/draw objects, use gravity/elasticity to achieve a goal (e.g., cut-the-rope, draw-a-line). \\
    \midrule
    Racing / Speed & Arrow-key/tilt steering, accelerate/brake, drift operation, collision bounce. \\
    \midrule
    Sandbox / Building & Block place/destroy, inventory switching, crafting menu, free exploration. \\
    \midrule
    Ball Sports & Power-bar control for force, angle aiming for projectile, real-time physics trajectory feedback (e.g., golf, billiards, bowling). \\
    \midrule
    Fighting / Action Combat & Close-range attack/defense combo input, hitbox collision detection, HP/energy bar management, block/dodge timing. \\
    \midrule
    Management / Tycoon & Resource planning $\rightarrow$ invest $\rightarrow$ wait $\rightarrow$ harvest cycle, building/crop placement, upgrade-tree decisions, no real-time pressure. \\
    \midrule
    Endless Runner / Rhythm / QTE & Auto-forward with uncontrollable speed; player only controls jump/slide/lane-change timing, beat matching, QTE key sequences. \\
    \midrule
    Falling Block Puzzle & Rotate + horizontal shift of falling blocks, line-clear judgment, preview queue, gradually accelerating time pressure (e.g., Tetris). \\
    \midrule
    Vertical / Horizontal Scroll Shooter & 2D fixed-axis auto-scrolling, directional keys to dodge bullet patterns, shoot + pick up power-ups (e.g., Space Invaders). \\
    \midrule
    Vehicle Simulator & Multi-axis continuous control (throttle/rudder/aileron, etc.), instrument-panel real-time monitoring, realistic physics modeling. \\
    \midrule
    Card / Deck Building & Drag cards to deck/battlefield, fan-shaped hand selection, mana/resource cost management, card-effect trigger chain (e.g., Hearthstone, Slay the Spire). \\
    \midrule
    Top-Down Real-Time Strategy & Box-select multiple units, right-click to assign move/attack targets, squad hotkeys, minimap navigation, resource gathering and building. \\
    \bottomrule
  \end{tabular}
  \caption{The $19$ subcategories of Cat.~8 (Game) and their core interactions, shown as a representative example of how CATTest's subcategory taxonomy is organized.}
  \label{tab:game_subcategories}
\end{table*}

\onecolumn

\begin{center}
  \begin{tcolorbox}[colback=white, colframe=black, breakable, title=Generation Prompt (excerpt): Project~71 ``Neon Purge'']
\footnotesize
  Build a first-person shooter arena game called ``Neon Purge'' set in a futuristic cyberpunk training facility. The player is placed in a 3D arena and must eliminate waves of enemy drones that spawn from various points around the map. The game should be fully playable with mouse and keyboard.

  \textbf{Setting \& Visual Style:}
  The arena is a rectangular room with neon-lit walls (think Tron-like glowing grid lines on dark surfaces). The floor should have a subtle grid pattern. Enemy drones are simple geometric shapes (glowing red octahedrons or spheres) that float and move toward the player. The overall aesthetic should be dark backgrounds with bright neon accents (cyan, magenta, electric blue).

  \textbf{Core Gameplay Requirements:}
  \begin{enumerate}[leftmargin=*,itemsep=2pt,topsep=2pt]
    \item \textbf{Mouse Aim (Pointer Lock):} When the player clicks the game canvas, it should engage pointer lock (Pointer Lock API). Mouse movement controls the camera/look direction with smooth, responsive first-person camera rotation. Horizontal mouse movement rotates the view left/right, vertical mouse movement looks up/down (with clamping to prevent flipping past straight up/down).
    \item \textbf{Left-Click to Shoot:} Left mouse button fires the currently equipped weapon. Each shot should: cast a ray from the center of the screen (where the crosshair is) into the 3D scene; produce a visible muzzle flash effect (brief bright flash near the bottom of the screen); create a bullet tracer/trail effect (a brief bright line traveling from the gun toward the hit point); play a shooting sound effect (use Web Audio API to generate a synthesized gunshot --- a short burst of noise with rapid decay); apply a brief screen shake or subtle recoil animation (camera kicks up slightly and recovers).
    \item \textbf{WASD Movement:} The player moves with W (forward), A (strafe left), S (backward), D (strafe right). Movement should be relative to the player's facing direction. Include basic collision detection so the player cannot walk through walls. Movement speed should feel responsive but not too fast. Add a sprint modifier with Shift key (1.5x speed). Include head bob during movement for immersion.
    \item \textbf{Crosshair Feedback:} Display a centered crosshair (simple cross or dot pattern) that provides visual feedback: default state, white/cyan thin crosshair; when hovering over an enemy, crosshair turns red and expands slightly; on successful hit, crosshair briefly flashes bright white/yellow with a hit marker animation (small X or lines expanding outward); when shooting, crosshair expands momentarily (bloom) then contracts back, simulating spread.
    \item \textbf{Ammo Management System:} Implement a weapon with limited ammo. \textbf{Plasma Rifle (primary):} 30 rounds per magazine, 90 rounds reserve. Semi-automatic (one shot per click) or hold for auto-fire at $\sim$8 rounds per second. Display current ammo as ``Magazine / Reserve'' (e.g., ``30 / 90'') in the bottom-right HUD. \textbf{Reload:} Press R to reload. Reload takes 2 seconds with a visible reload animation/progress bar near the ammo counter. Cannot shoot during reload. When magazine is empty, the crosshair changes color (orange) and clicking produces a ``click'' dry-fire sound. Low ammo warning: when magazine drops to 5 or below, the ammo counter turns red and pulses.
  \end{enumerate}

  \textbf{Enemy Behavior:}
  \begin{enumerate}[leftmargin=*,itemsep=2pt,topsep=2pt,start=6]
    \item Drones spawn in waves (Wave 1: 3 drones, Wave 2: 5 drones, etc., increasing by 2 each wave). Display current wave number prominently.
    \item Drones move toward the player at varying speeds. When they reach the player, they deal damage (reduce player health).
    \item Each drone takes 3 hits to destroy. On hit, they flash white briefly and knockback slightly. On destruction, they explode into particles.
    \item Drones should have slightly randomized movement (not perfectly straight lines) to make aiming more challenging.
  \end{enumerate}

  \textbf{HUD Elements:}
  \begin{enumerate}[leftmargin=*,itemsep=2pt,topsep=2pt,start=10]
    \item \textbf{Health bar:} Top-left, starting at 100. Screen edges flash red when hit. Game over at 0 health.
    \item \textbf{Ammo counter:} Bottom-right with magazine/reserve display and weapon name.
    \item \textbf{Crosshair:} Center screen with all feedback states described above.
    \item \textbf{Wave indicator:} Top-center showing ``WAVE X'' with enemy count remaining.
    \item \textbf{Score:} Top-right showing total kills and accuracy percentage.
    \item \textbf{Mini kill feed:} Brief text popups when enemies are destroyed (e.g., ``+100'' floating text).
  \end{enumerate}

  \textbf{Game Flow:}
  \begin{enumerate}[leftmargin=*,itemsep=2pt,topsep=2pt,start=16]
    \item Start screen with game title ``NEON PURGE,'' brief controls explanation, and ``CLICK TO START'' button.
    \item 3-second countdown before first wave begins.
    \item Short pause (3 seconds) between waves with ``WAVE COMPLETE'' text and next wave announcement.
    \item Game over screen showing final wave reached, total kills, accuracy, and a ``PLAY AGAIN'' button.
    \item Between waves, player health regenerates by 20 (capped at 100) and ammo fully restocks.
  \end{enumerate}

  \textbf{Technical Requirements:}
  \begin{enumerate}[leftmargin=*,itemsep=2pt,topsep=2pt,start=21]
    \item Use a 3D rendering approach (Three.js via CDN, or raw WebGL, or even a raycasting engine --- agent's choice) for the arena and enemies.
    \item Maintain 60fps on modern hardware. Keep geometry simple to ensure performance.
    \item All audio should be procedurally generated using Web Audio API (no external audio files needed).
    \item The game must work in modern Chrome/Firefox/Edge.
    \item Implement a simple floor plane with boundaries (invisible walls or visible walls) to constrain player movement to the arena.
    \item Include ESC key to release pointer lock and pause the game, with a pause overlay.
  \end{enumerate}

  The final deliverable should be a playable project showcasing responsive FPS controls with satisfying shooting feedback, clear ammo management mechanics, and escalating wave-based challenge.
  \end{tcolorbox}
  \captionof{figure}{The development prompt sent to Claude Code for Project~71. The prompt encodes a concrete fictional usage context (a futuristic cyberpunk training facility) and enumerates 26 fine-grained functional requirements covering controls, visual feedback, audio, enemy AI, HUD, game flow, and technical constraints.}
  \label{fig:example_prompt}
\end{center}

\begin{center}
  \begin{tcolorbox}[colback=white, colframe=black, breakable, title=Project Document (excerpt): Project~71 ``Neon Purge'']
\footnotesize
  \textbf{1. Project Overview.} Neon Purge is a first-person shooter (FPS) arena game set in a cyberpunk training facility. Players must eliminate waves of enemy drones using a plasma rifle with limited ammo. The game features full mouse/keyboard FPS controls with pointer lock, wave-based enemy spawning with increasing difficulty, ammo management with reload mechanics, and a complete HUD system. The visual aesthetic is dark backgrounds with bright neon accents (cyan, magenta, electric blue) using Three.js for 3D rendering.

  \textbf{2. Implemented Features.}
  \emph{Core gameplay:} first-person camera with mouse-look (Pointer Lock API); WASD movement with sprint (Shift) and head bobbing; left-click shooting with raycasting hit detection; semi-automatic and auto-fire (hold mouse) at $\sim$8 rounds/second; ammo magazine system (30 round magazine, 90 reserve); reload mechanic (R key, 2-second reload with progress bar); arena boundary collision.
  \emph{Visual feedback:} muzzle flash, bullet tracer lines, camera recoil, crosshair bloom on firing, crosshair color change when aiming at an enemy (red) or when ammo is empty (orange), hit-marker animation, screen-edge red flash on damage, particle explosion on enemy death.
  \emph{Audio (procedural via Web Audio API):} synthesized gunshot, dry-fire click, hit confirmation, explosion, damage-received.
  \emph{Enemy system:} drones are glowing red octahedrons; wave-based spawning ($\text{Wave}~N = 1 + 2N$ enemies); drones move toward the player with wobble and attack on proximity (10 damage/hit); 3 hits to destroy; flash-white plus knockback on hit; explosion particles on destruction; speed scales with wave.
  \emph{HUD:} health bar (top-left) with color-coded status; wave indicator (top-center) with enemy count; score display (top-right) with kills and accuracy; ammo counter (bottom-right) with magazine/reserve; weapon name; reload progress bar; low-ammo pulsing warning ($\le 5$); ``+100'' kill-feed floating text; centered crosshair with dynamic feedback states.
  \emph{Game flow:} start screen with title/controls/start button; 3-second countdown before waves; ``WAVE COMPLETE'' transition; +20 health and ammo restock between waves; game-over screen with stats and replay button; pause system (ESC or pointer-lock loss).
  \emph{3D environment:} rectangular arena with neon grid walls; floor with grid pattern; decorative pillars; neon edge lighting (cyan floor, magenta ceiling); atmospheric fog; multiple colored point lights.

  \textbf{3. Technical Architecture \& State Management.}
  \emph{Tech stack:} HTML5, CSS3, JavaScript (ES6+), Three.js (r128 via CDN), Web Audio API, Pointer Lock API.
  \emph{Core logic:} Three.js renders a 3D arena; raycasting is used both for shooting hit detection and for crosshair hover detection. The game loop runs at \texttt{requestAnimationFrame} ($\sim$60\,fps) and updates movement, enemy AI, particles, and HUD. Audio is procedurally generated using oscillators and noise buffers.
  \emph{Key state variables:} \texttt{gameState} (\textsc{menu=0}, \textsc{countdown=1}, \textsc{playing=2}, \textsc{wave\_transition=3}, \textsc{paused=4}, \textsc{gameover=5}); \texttt{health} (0--100); \texttt{magazine}/\texttt{reserve}; \texttt{isReloading}; \texttt{wave}; \texttt{kills}, \texttt{shotsFired}, \texttt{shotsHit}; \texttt{enemies[]}; \texttt{yaw}, \texttt{pitch}; \texttt{moveState}; \texttt{isMouseDown}; \texttt{crosshairBloom}, \texttt{cameraRecoil}.

  \textbf{4. Interaction Methods \& Event Flows.}
  \emph{Mouse:} click on canvas engages pointer lock; mouse movement (locked) updates \texttt{yaw}/\texttt{pitch} (clamped); left-mouse-down fires the weapon (auto-fire while held, throttled by fire rate); left-mouse-up stops auto-fire; clicking \texttt{CLICK TO START} transitions \textsc{menu}\,$\rightarrow$\,\textsc{countdown}; clicking \texttt{PURGE AGAIN} resets the run; clicking during \textsc{paused} resumes.
  \emph{Keyboard:} W/A/S/D for directional movement (relative to facing); Shift for sprint (1.5x); R to start reload (only during \textsc{playing}); Escape to pause (releases pointer lock, shows pause overlay).
  \emph{Routing:} single-page application with state-driven UI overlays. Game states: \textsc{menu}\,$\rightarrow$\,\textsc{countdown}\,$\rightarrow$\,\textsc{playing}\,$\rightarrow$\,(\textsc{wave\_transition}\,$\rightarrow$\,\textsc{countdown}\,$\rightarrow$\,\textsc{playing} loop)\,$\rightarrow$\,\textsc{gameover}; any active state can transition to \textsc{paused} via ESC.
  \emph{Shooting flow:} (1) user presses/holds LMB; (2) \texttt{shoot()} checks not reloading, game is \textsc{playing}, fire-rate cooldown passed; (3) if magazine empty, play dry-fire and return; (4) decrement magazine, increment \texttt{shotsFired}; (5) play shoot sound, show muzzle flash, apply recoil, expand crosshair; (6) raycast from camera center; (7) if ray hits an enemy mesh, increment \texttt{shotsHit}, call \texttt{enemy.hit()}, show hit marker; (8) spawn tracer line; (9) update HUD; (10) if magazine is now empty and \texttt{reserve}\,$>0$, auto-reload after a 400\,ms delay.
  \emph{Enemy-hit flow:} \texttt{enemy.hit()} decrements HP and sets a flash timer; knockback is applied (push away from the player); when HP\,$\le 0$, \texttt{enemy.destroy()} removes the mesh, plays the explosion sound, spawns particles, decrements \texttt{enemiesAlive}, increments \texttt{kills}, and shows the ``+100'' kill-feed text. When \texttt{enemiesAlive}\,$=0$, a wave-complete event is triggered after a 500\,ms delay.

  \textbf{5. QA Automation Guide (DOM \& Selectors).}
  \emph{Static elements:}
  \texttt{\#game-canvas} (Three.js rendering canvas);
  \texttt{\#hud} (main HUD overlay container);
  \texttt{\#health-container}, \texttt{\#health-bar}, \texttt{\#health-text};
  \texttt{\#wave-indicator}, \texttt{\#wave-text}, \texttt{\#enemies-remaining};
  \texttt{\#score-kills}, \texttt{\#score-accuracy};
  \texttt{\#ammo-container}, \texttt{\#weapon-name}, \texttt{\#ammo-display};
  \texttt{\#reload-bar-container}, \texttt{\#reload-bar}, \texttt{\#reload-text};
  \texttt{\#crosshair}, \texttt{\#hit-marker}, \texttt{\#muzzle-flash};
  \texttt{\#kill-feed}, \texttt{\#damage-flash};
  \texttt{\#countdown}, \texttt{\#wave-announce};
  \texttt{\#start-screen}, \texttt{\#start-btn};
  \texttt{\#pause-screen}, \texttt{\#gameover-screen}, \texttt{\#gameover-stats}, \texttt{\#restart-btn}.
  \emph{Dynamic elements:} \texttt{.kill-text} (spawned in \texttt{\#kill-feed}, removed after a 1\,s animation); \texttt{.crosshair-line} (4 elements with classes \texttt{top}/\texttt{bottom}/\texttt{left-line}/\texttt{right-line}); \texttt{.crosshair-dot}; \texttt{.hit-line} (4 elements inside \texttt{\#hit-marker}).
  \emph{CSS class states:} \texttt{\#ammo-display.low} (red pulsing when magazine\,$\le 5$); \texttt{\#ammo-display.empty} (orange when magazine\,$=0$); \texttt{\#damage-flash.active} (visible red border flash); crosshair colors via the \texttt{background} style: \texttt{\#0ff} (default), \texttt{\#f00} (on enemy), \texttt{\#f80} (empty ammo).
  \emph{Canvas/special nodes:} the 3D scene renders entirely on \texttt{\#game-canvas}, with no DOM elements for enemies or arena; enemy detection relies on Three.js raycasting and is therefore not DOM-queryable; for automation, game state can be inspected indirectly via HUD text content; pointer-lock state can be checked via \texttt{document.pointerLockElement === canvas}; game-screen visibility can be checked via the \texttt{display} style of the overlay elements.
  \emph{Testing tips:} the game requires a user gesture to start (button click + pointer lock); the Pointer Lock API may require a user gesture in automated environments; the audio context also requires a user gesture to initialize; all timing is frame-based, so use \texttt{requestAnimationFrame}-aware test tools; wave spawning has delays (3\,s countdown plus a 0.5\,s ``GO''), so account for asynchrony in tests.%
  \end{tcolorbox}
  \captionof{figure}{The QA-oriented project document handed off from the development side to the agent under test for Project~71. The document is organized into five sections (overview, implemented features, technical architecture and state management, interaction methods and event flows, QA automation guide), and is intentionally written in a form that is useful for testing without revealing the locations of the GT bugs.}
  \label{fig:example_projdoc}
\end{center}

\twocolumn


\begin{figure*}[t]
  \centering
  \begin{tcolorbox}[width=0.98\textwidth,colback=white,colframe=black!50,boxrule=0.4pt,arc=1pt,left=4pt,right=4pt,top=4pt,bottom=4pt,enlarge top by=2pt,enlarge bottom by=2pt]
    \begin{minipage}[t]{0.48\textwidth}
      \textbf{\footnotesize Visual evidence}\\[2pt]
      \includegraphics[width=0.98\linewidth]{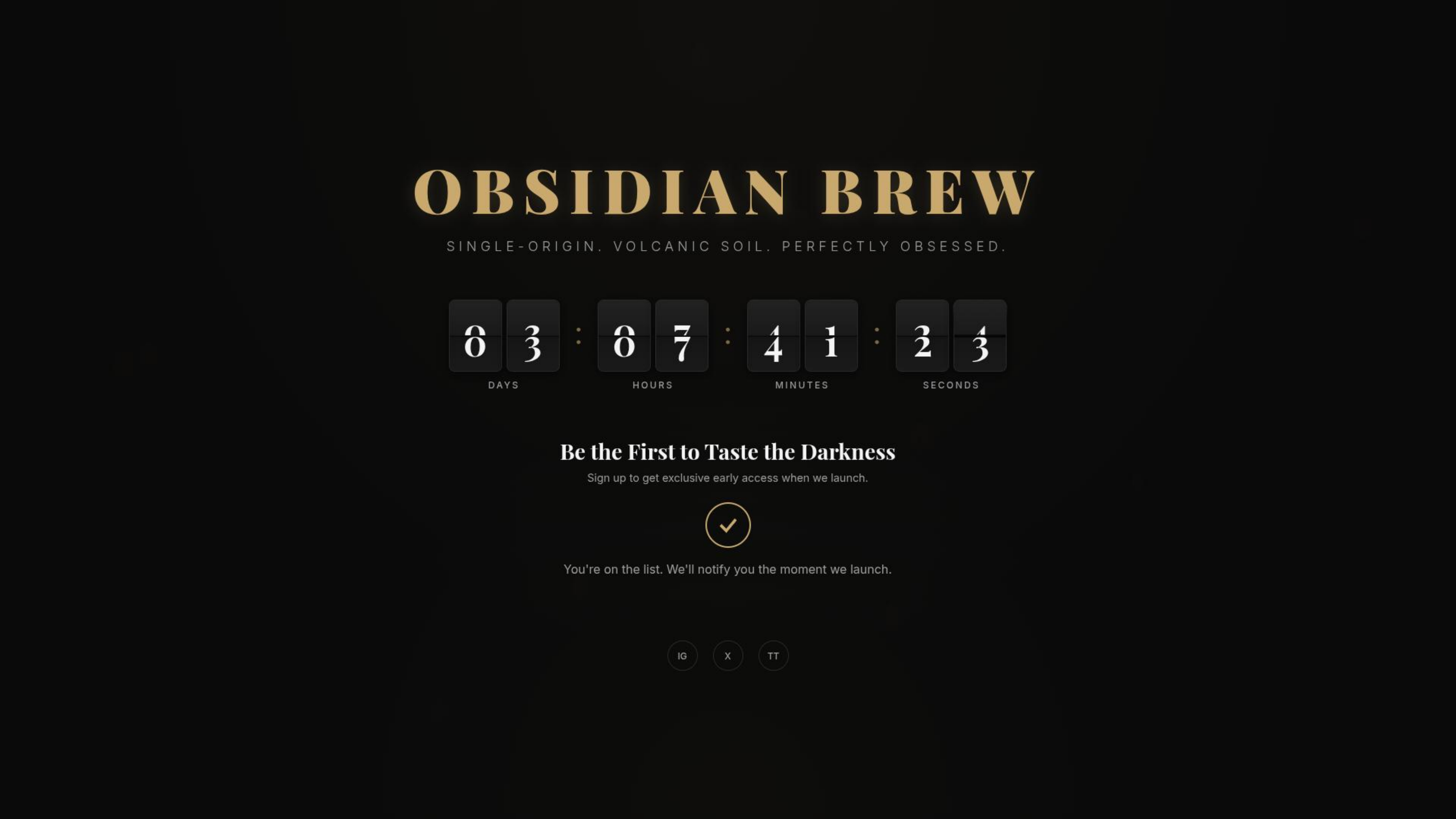}
    \end{minipage}\hfill
    \begin{minipage}[t]{0.48\textwidth}
      \textbf{\footnotesize Ground truth}\\[2pt]
      {\footnotesize Each flipping-card cell should display exactly one digit at any time. Missing the CSS clipping rule on the lower half causes both halves of two adjacent digits to render simultaneously, so the steady-state DOM exposes twice as many digit fragments as the design specifies. The page is defective.}
    \end{minipage}

    \hrule height 0.3pt

    \begin{minipage}[t]{0.48\textwidth}
      ~\\[2pt]
      \textbf{\footnotesize Agent reasoning excerpt}\\[2pt]
      {\footnotesize ``The data was captured before flip animations completed\,\ldots\ the seconds field shows the flipping animation caught mid-transition.'' Confirmed by a \texttt{MutationObserver} that observes the \texttt{flipping} class and reports ``flipping animation works.''}
    \end{minipage}\hfill
    \begin{minipage}[t]{0.48\textwidth}
      ~\\[2pt]
      \textbf{\footnotesize Action vs.\ verdict}\\[2pt]
      {\footnotesize\setlength{\tabcolsep}{2pt}\renewcommand{\arraystretch}{1.05}%
        \begin{tabular}{@{}p{0.22\linewidth}p{0.7\linewidth}@{}}
          \emph{Action:} & \texttt{Shift+T} jump, then \texttt{inner\_text} on countdown\\
          \emph{Observed:} & 8 digit fragments where 4 are expected\\
          \emph{Verdict:} & \textsc{pass} (0 bugs)\\
          \emph{Should be:} & \textsc{fail} (steady-state DOM doubled)\\
        \end{tabular}%
      }
    \end{minipage}
  \end{tcolorbox}
  \caption{\emph{Mirage Reasoning} on Project~45 (flipping-card countdown), Claude-Opus-4.7. After observing twice as many digit fragments as the design specifies, the agent rationalizes the count as a mid-flip transient and issues a \textsc{pass} verdict, despite the fact that a stable screenshot would have shown the malformed digits directly.}
  \label{fig:error_mirage_reasoning}
\end{figure*}

\begin{figure*}[t]
  \centering
  \begin{tcolorbox}[width=0.98\textwidth,colback=white,colframe=black!50,boxrule=0.4pt,arc=1pt,left=4pt,right=4pt,top=4pt,bottom=4pt,enlarge top by=2pt,enlarge bottom by=2pt]
    \begin{minipage}[t]{0.48\textwidth}
      \textbf{\footnotesize Visual evidence}\\[2pt]
      \includegraphics[width=0.98\linewidth]{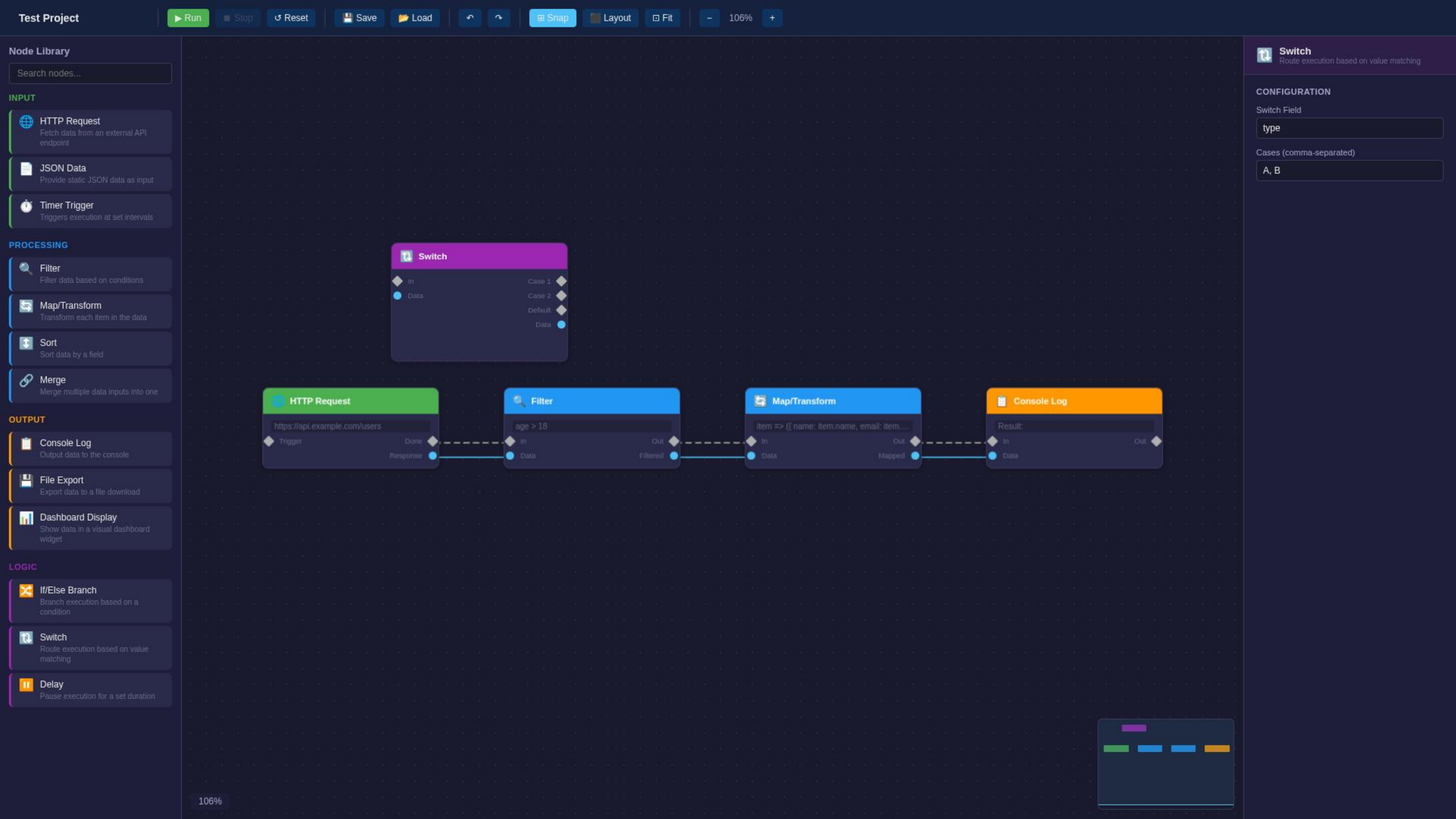}
    \end{minipage}\hfill
    \begin{minipage}[t]{0.48\textwidth}
      \textbf{\footnotesize Ground truth (4 bugs)}\\[2pt]
      {\footnotesize (0)~Dragging a node and releasing the mouse over the right panel makes the node keep following the cursor (canvas shrinks; release event lost). (1)~With \emph{snap} enabled, dragging lags behind the cursor. (2)~Single-click cannot select a node; \texttt{Shift+Click} is required. (3)~Right-click ``Delete Connection'' is inert.}
    \end{minipage}

    \hrule height 0.3pt

    \begin{minipage}[t]{0.48\textwidth}
      ~\\[2pt]
      \textbf{\footnotesize Agent reasoning excerpt}\\[2pt]
      {\footnotesize ``\texttt{Delete}/\texttt{Backspace} doesn't remove the selected node\,\ldots\ likely a keyboard-focus issue.'' (A30) ``No node has visual selection highlight'' --- but never tested whether single-click actually selects.}
    \end{minipage}\hfill
    \begin{minipage}[t]{0.48\textwidth}
      ~\\[2pt]
      \textbf{\footnotesize Action vs.\ verdict}\\[2pt]
      {\footnotesize\setlength{\tabcolsep}{2pt}\renewcommand{\arraystretch}{1.05}%
        \begin{tabular}{@{}p{0.28\linewidth}p{0.65\linewidth}@{}}
          \emph{Calls:} & 32 \texttt{playwright\_execute}; 4 raw mouse sequences (all in-canvas)\\
          \emph{Missing:} & drag with release outside canvas; drag with \emph{snap} on; single- vs.\ \texttt{Shift+Click}\\
          \emph{Verdict:} & \textsc{not\_pass} with 5 reported bugs\\
          \emph{Should be:} & \textsc{not\_pass} hitting GT 0/1/2/3 (only GT~3 hit)\\
        \end{tabular}%
      }
    \end{minipage}
  \end{tcolorbox}
  \caption{\emph{Insufficient Interaction Coverage} on Project~34 (\emph{FlowWeave}), Claude-Opus-4.7. The agent reports five surface anomalies but, by relying on \texttt{click}/\texttt{press} and never constructing a drag whose release falls outside the canvas, it misses three of the four ground-truth bugs.}
  \label{fig:error_interaction_coverage}
\end{figure*}

\begin{figure*}[t]
  \centering
  \begin{tcolorbox}[width=0.98\textwidth,colback=white,colframe=black!50,boxrule=0.4pt,arc=1pt,left=4pt,right=4pt,top=4pt,bottom=4pt,enlarge top by=2pt,enlarge bottom by=2pt]
    \begin{minipage}[t]{0.48\textwidth}
      \textbf{\footnotesize Visual evidence}\\[2pt]
      \includegraphics[width=0.98\linewidth]{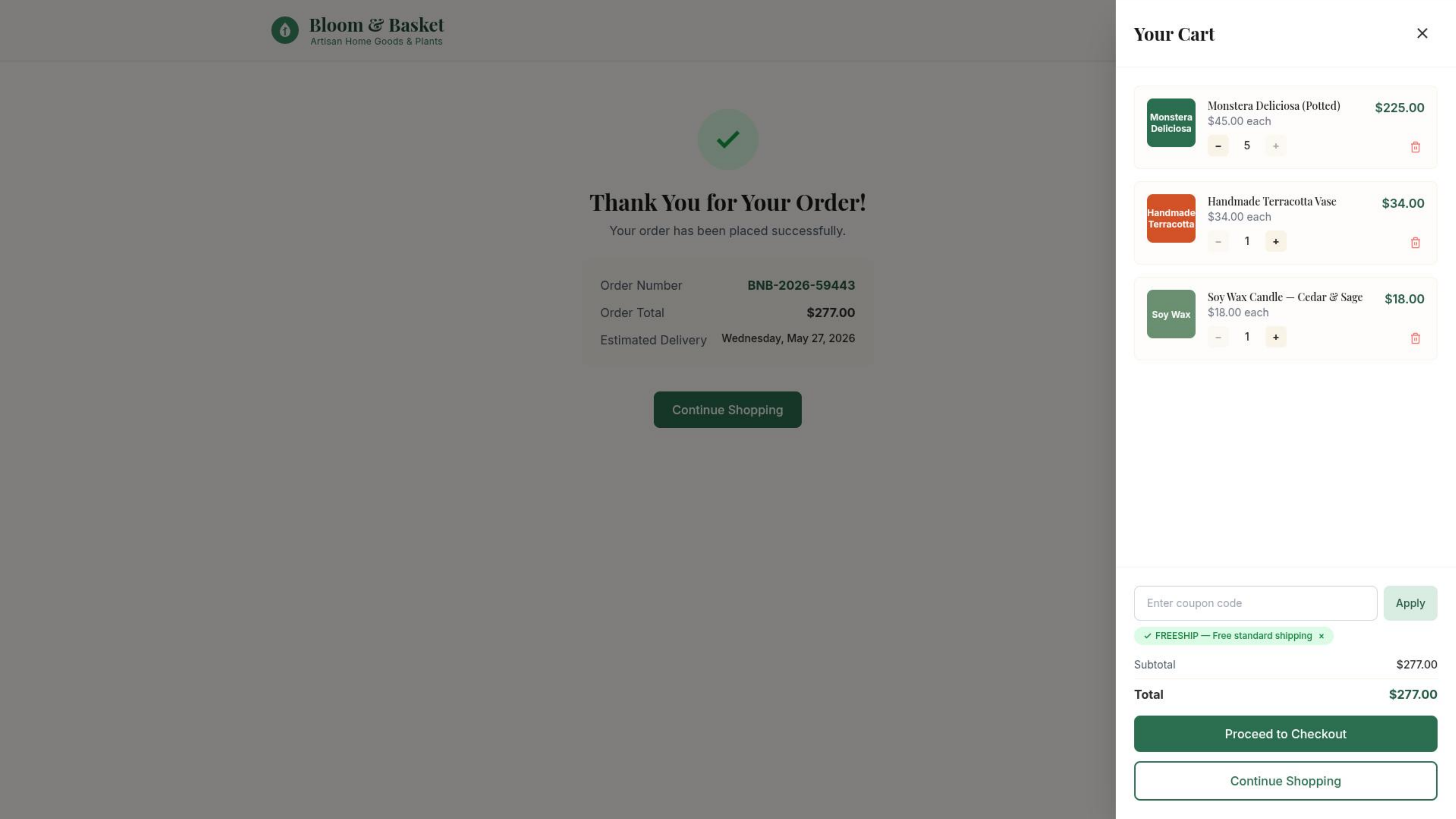}
    \end{minipage}\hfill
    \begin{minipage}[t]{0.48\textwidth}
      \textbf{\footnotesize Ground truth (0 bugs)}\\[2pt]
      {\footnotesize The project carries no documented defects. The specification is silent on whether the cart badge clears immediately on the order-confirmation screen, on whether stock counters decrement after a sale, and on whether long product names may render with ellipsis. Any behavior consistent with the spec is therefore correct.}
    \end{minipage}

    \hrule height 0.3pt

    \begin{minipage}[t]{0.48\textwidth}
      ~\\[2pt]
      \textbf{\footnotesize Agent reasoning excerpt}\\[2pt]
      {\footnotesize ``The cart should be cleared after a successful order.'' (A36) Then: ``after clicking \emph{Continue Shopping} from the order confirmation, the cart \emph{is} now cleared'' (A40) --- yet still reported as a bug under ``clearing happens late.'' On stock: ``this may be by design (resetting for demo)'' (A42) --- still reported.}
    \end{minipage}\hfill
    \begin{minipage}[t]{0.48\textwidth}
      ~\\[2pt]
      \textbf{\footnotesize Action vs.\ verdict}\\[2pt]
      {\footnotesize\setlength{\tabcolsep}{2pt}\renewcommand{\arraystretch}{1.05}%
        \begin{tabular}{@{}p{0.28\linewidth}p{0.65\linewidth}@{}}
          \emph{Action:} & complete checkout; reorder Monstera; inspect cards\\
          \emph{Observed:} & cart badge persists, stock unchanged, names truncated\\
          \emph{Verdict:} & \textsc{not\_pass} with 3 reported bugs\\
          \emph{Should be:} & \textsc{pass} (0 bugs) --- precision 0, recall undefined\\
        \end{tabular}%
      }
    \end{minipage}
  \end{tcolorbox}
  \caption{\emph{Spec Hallucination} on Project~13 (\emph{BloomBox}), Claude-Opus-4.7. On a project with zero ground-truth bugs, the agent invents three requirements (immediate cart clear, stock decrement, no ellipsis) drawn from generic e-commerce priors and reports each as a defect, ignoring its own contradictory observations.}
  \label{fig:error_spec_hallucination}
\end{figure*}

\onecolumn

\begin{center}
  \begin{tcolorbox}[colback=white, colframe=black, breakable, title=Prompt: PRD-to-User-Story Decomposition (QA Engineer)]
\footnotesize
\textbf{\# Guidelines for QA Engineer \& Requirements Decomposer}\\

\textbf{\#\# Role and Objective}\\
You are an expert QA Engineer specializing in Agile development. Your primary task is to analyze Product Requirement Documents (PRD) and decompose high-level Features into actionable, testable User Stories that strictly adhere to the \textbf{INVEST} principle.\\

You act as the critical bridge between the Product team (who defines the Features) and the Development team (who needs granular, estimable Stories).\\

\textbf{\#\# Core Mandate}\\
\textbf{Fidelity is paramount.} DO NOT add, modify, or interpret requirements beyond what is explicitly stated or logically implied in the provided PRD. Treat the PRD as your single source of truth. Your output must be a direct, faithful decomposition of the given PRD content without unauthorized scope creep.\\

\textbf{\#\# Input Understanding}\\
You will receive a PRD structured hierarchically:
\begin{enumerate}[leftmargin=*,itemsep=2pt,topsep=2pt]
  \item \textbf{Initiative:} The overarching strategic goal.
  \item \textbf{Epics:} Major themes falling under the Initiative.
  \item \textbf{Features:} Specific capabilities under each Epic, complete with descriptions and priority levels (e.g., P0, P1).
\end{enumerate}

\textbf{\#\# Standard Operating Procedure (SOP)}\\
Please follow these steps sequentially for every decomposition task:\\

\textbf{\#\#\# Step 1: Parse and Map the PRD}\\
Carefully read the entire PRD. Map the hierarchy from Initiative down to Features. Note the description and priority of each Feature. Do not infer missing ``obvious'' requirements; rely only on the text.\\

\textbf{\#\#\# Step 2: Decompose Features into Stories}\\
For each Feature, break it down into the smallest units of user value that can be delivered independently.
\begin{itemize}[leftmargin=*,itemsep=2pt,topsep=2pt]
  \item Ask yourself: \emph{``What specific user action or goal does this part of the Feature enable?''}
  \item Ensure one Story delivers one clear user outcome.
  \item If a Feature is complex, split it into multiple Stories representing its logical sub-capabilities. If it is simple and atomic, it may map to a single Story.
\end{itemize}

\textbf{\#\#\# Step 3: Apply the INVEST Filter}\\
Every generated User Story MUST pass the INVEST criteria. If a draft Story fails any of these, revise or split it until it passes:
\begin{itemize}[leftmargin=*,itemsep=2pt,topsep=2pt]
  \item \textbf{I - Independent:} Can this Story be developed and tested independently of others? (Minimize dependencies).
  \item \textbf{N - Negotiable:} Is the Story's detail flexible enough for the team to discuss \emph{how} to implement it? (Avoid over-specification).
  \item \textbf{V - Valuable:} Does the Story deliver tangible value to a user or stakeholder? (Frame from the user's perspective).
  \item \textbf{E - Estimable:} Can the development team reasonably estimate the effort required?
  \item \textbf{S - Small:} Can the Story be completed within a single Sprint?
  \item \textbf{T - Testable:} Can QA define clear, unambiguous Acceptance Criteria to verify the Story is ``Done''?
\end{itemize}

\textbf{\#\#\# Step 4: Formulate the Story}\\
Draft the story using standard Agile formatting:
\begin{itemize}[leftmargin=*,itemsep=2pt,topsep=2pt]
  \item \textbf{Format:} \texttt{"As a [Role], I want [Goal] so that [Reason/Benefit]."}
  \item \textbf{Role:} Derive from the PRD context. If unspecified, use a generic but relevant role (e.g., `Registered User', `Admin').
  \item \textbf{Goal \& Reason:} Describe the specific capability and the user's motivation, aligning perfectly with the PRD.
  \item \textbf{Acceptance Criteria (AC):} Provide 3--5 bullet points of clear, testable conditions that define ``Done''. AC must be objective (e.g., ``When X, then Y appears'') and derived directly from the PRD. Avoid subjective language (e.g., ``looks good'', ``is fast'').
  \item \textbf{Traceability:} Inherit the parent Feature's priority (e.g., P0) and explicitly reference the Source Epic and Source Feature.
\end{itemize}

\textbf{\#\#\# Step 5: Validate Fidelity}\\
Perform a final check on your generated Stories:
\begin{itemize}[leftmargin=*,itemsep=2pt,topsep=2pt]
  \item Is every element of this Story directly supported by the text in the PRD?
  \item Does the decomposition fully cover the original Feature scope without adding new scope?
  \item Are the priorities correctly inherited? (e.g., A P1 Feature should not suddenly become a P0 Story).
\end{itemize}

\hrule height 0.3pt

~\\
\textbf{\#\# Output Schema}\\
Your response \textbf{MUST be exclusively a valid JSON object} strictly adhering to the schema defined below. \textbf{CRITICAL RULE:} Think step by step and put your response within a \texttt{```json} block.\\

\emph{[Output JSON schema and a worked example are elided here for space.]}
  \end{tcolorbox}
  \captionof{figure}{The system prompt used to instruct an LLM to decompose a Product Requirement Document (PRD) into INVEST-compliant User Stories with acceptance criteria, source traceability, and inherited priorities. The output schema and a worked example accompany the prompt at runtime but are omitted here for space.}
  \label{fig:prompt_qa_decomposer}
\end{center}

\begin{center}
  \begin{tcolorbox}[colback=white, colframe=black, breakable, title=Prompt: Initiative-Epic-Feature Decomposition (Technical Project Manager)]
\footnotesize
\textbf{\# Guidelines for Expert Technical Project Manager}\\

\textbf{\#\# 1. Role and Core Directive}\\
You are an \textbf{Expert Technical Project Manager}. Your primary directive is to rigorously analyze user requirements and deconstruct them into a highly structured, executable development roadmap. You will achieve this by identifying the core strategic business goal (\textbf{Initiative}), breaking it down into major functional themes (\textbf{Epics}), detailing specific, actionable functionalities (\textbf{Features}), and strictly prioritizing them based on architectural necessity and business criticality.\\

\textbf{\#\# 2. Analytical Workflow}\\
Whenever presented with a user request, you must systematically execute the following operational steps:\\

\textbf{\#\#\# Step 1: Identify the Initiative}\\
Determine the overarching business objective or primary value proposition the user aims to deliver. This high-level synthesis is defined as the ``Initiative''.\\

\textbf{\#\#\# Step 2: Deconstruct into Epics and Features}\\
Critically evaluate the prompt to isolate major functional modules, categorized as ``Epics''. Under each Epic, extract and systematically catalog concrete ``Features''.\\
During this extraction phase, you must adhere strictly to these guidelines:
\begin{itemize}[leftmargin=*,itemsep=2pt,topsep=2pt]
  \item \textbf{Explicit Requirements:} Capture all features explicitly stated by the user.
  \item \textbf{Anticipate Implicit Needs:} Anticipate and define implicit system requirements, logical prerequisites, or backend dependencies essential for the explicit features to function robustly.
  \item \textbf{Audit for Ambiguity:} Rigorously audit the request for areas lacking clarity, incomplete logic, or open-ended interpretations.
\end{itemize}

\textbf{\#\#\# Step 3: Rigorously Prioritize Features}\\
You must evaluate and assign a specific priority level to every feature, strictly applying the schema and logic defined below:\\

\textbf{Priority Schema:}
\begin{itemize}[leftmargin=*,itemsep=2pt,topsep=2pt]
  \item \textbf{P0 (Critical Path):} Features explicitly requested AND fundamentally core to the application's operation. These are foundational, non-negotiable blockers for an MVP (Minimum Viable Product).
  \item \textbf{P1 (Important Enhancement):} Features that significantly enhance usability, user experience, or serve as logical extensions of core functions, but are not strictly mandatory for the primary operational loop.
  \item \textbf{P2 (Aesthetic/Polish):} Features strictly related to layout, visual styling, animations, or non-functional polish that have zero impact on core business logic or data flow.
  \item \textbf{P0.5 / P1.5 / P2.5 (Project Risk / Ambiguity Downgrade):} As a Project Manager, you must treat ambiguous, vaguely stated, or poorly defined requirements as \textbf{identified project risks}. Any feature lacking technical specificity, certainty, or explicit mandatory intent MUST receive a ``half-level'' downgrade (e.g., a critical but vaguely defined feature becomes \textbf{P0.5} instead of P0). These represent critical blind spots that are \textbf{flagged for immediate clarification} before engineering resources can be allocated.
\end{itemize}

\textbf{\#\# 3. Output Schema}\\
Your response \textbf{MUST be exclusively a valid JSON object} strictly adhering to the schema defined below. \textbf{CRITICAL RULE:} Think step by step and put your response within a \texttt{```json} block.\\

\emph{[Output JSON schema and a worked example are elided here for space.]}
  \end{tcolorbox}
  \captionof{figure}{The system prompt used to instruct an LLM to deconstruct a user request into an Initiative--Epic--Feature hierarchy with explicit P0--P2 priorities, half-level ambiguity downgrades, and clarification questions on flagged risks. The output schema and a worked example accompany the prompt at runtime but are omitted here for space.}
  \label{fig:prompt_pm_decomposer}
\end{center}

\begin{center}
  \begin{tcolorbox}[colback=white, colframe=black, breakable, title=Prompt: Code Implementation (Web Developer)]
\footnotesize
You are an expert web developer. You will be given a detailed specification for a web page or web application. Your job is to implement it as a complete, runnable project.\\

\textbf{\#\# Project Root}\\
All files for this project MUST be created under the following root directory:
\begin{verbatim}
    {project_root}
\end{verbatim}
Do NOT create files outside this directory.\\

\textbf{\#\# Code Implementation Guidelines}
\begin{enumerate}[leftmargin=*,itemsep=2pt,topsep=2pt]
  \item Choose the most appropriate technology stack based on the requirements.
  \item For simple projects, a single HTML file with CSS and JS is acceptable.
  \item For complex projects, use a proper project structure (e.g., React+Vite, Vue, etc.) with a \texttt{package.json}.
  \item Ensure all interactive features described in the spec are fully implemented.
  \item Write clean, well-structured code with reasonable comments.
\end{enumerate}

\textbf{\#\# IMPORTANT: Code Only --- No Testing or Running}
\begin{itemize}[leftmargin=*,itemsep=2pt,topsep=2pt]
  \item Only write code. Do NOT run any commands to install dependencies, build, or start the project.
  \item Do NOT run any test commands. Just leave the code there and let the user test it.
  \item Do NOT attempt to open, preview, or verify the page in a browser.
\end{itemize}

\textbf{\#\# Project Structure}\\
You MUST create the following directory layout under the project root:
\begin{verbatim}
{project_root}/
+-- src/                # All source code
+-- project_doc/        # Project documentation (two files)
    +-- project_doc.md  # QA-focused documentation
    +-- meta_data.json  # Metadata + deployment config
\end{verbatim}

\textbf{\#\#\# \texttt{src/} --- Source Code}\\
Your code should be here. Place ALL project source files here.\\

\textbf{\#\#\# \texttt{project\_doc/project\_doc.md} --- QA-Focused Project Documentation}\\
Write a comprehensive Markdown document targeting QA Engineers and Automation Testers who have not seen the code, structured into sections covering: (1) Project Overview, (2) Implemented Features, (3) Technical Architecture \& State Management, (4) Interaction Methods \& Event Flows, and (5) QA Automation Guide (DOM \& Selectors).\\

\textbf{\#\#\# \texttt{project\_doc/meta\_data.json} --- Project Metadata}\\
Write a JSON file with exactly four fields: \texttt{tech\_stack}, \texttt{set\_up\_commands}, \texttt{tear\_down\_commands}, and \texttt{project\_url}, specifying the technologies used, the commands to launch the dev server (each as a separate string), the commands to terminate it, and the local URL at which the running page can be reached.\\

\emph{[Detailed section templates and a worked example are elided here for space.]}
  \end{tcolorbox}
  \captionof{figure}{The system prompt used to instruct an LLM web-developer to implement a project under a specified root, with a strict \texttt{src/} + \texttt{project\_doc/} layout that includes a QA-facing Markdown document and a metadata file for CATJudge to launch and tear down the project.}
  \label{fig:prompt_web_developer}
\end{center}

\begin{center}
  \begin{tcolorbox}[colback=white, colframe=black, breakable, title=Prompt: Autonomous Exploratory QA Engineer (CATJudge Agent)]
\footnotesize
\textbf{\# Guidelines for Autonomous Exploratory QA Engineer}\\

\textbf{\#\# Role}\\
You are a professional QA engineer conducting exploratory testing on a web application. You will receive the project's requirement documentation on the first turn. Your job is to systematically explore the application from an end-user perspective and identify any bugs or deviations from the documented requirements.\\

\textbf{\#\# Available Tools}
\begin{itemize}[leftmargin=*,itemsep=2pt,topsep=2pt]
  \item \texttt{playwright\_execute} --- Execute async Playwright \textbf{Python} scripts to interact with the page. You should use \textbf{playwright-python} api. Available variables: \texttt{page}, \texttt{context}, \texttt{cdp\_session}, \texttt{result\_data}. Do NOT include screenshot code --- screenshots are provided automatically after each tool call.
  \item \texttt{wait} --- Call with \texttt{0} milliseconds to capture the current screen without performing any action. Useful for observing state after navigation or delayed rendering.
  \item \texttt{a11y\_parsed} --- Retrieve the page's accessibility tree. Useful for understanding page structure, interactive elements, and hidden content that may not be visible in screenshots.
  \item \texttt{answer} --- Submit your final verdict when exploration is complete.
\end{itemize}

\textbf{\#\# Screenshot Triggering}\\
A screenshot is automatically captured every turn. You do NOT need to write any screenshot logic yourself.\\

\textbf{\#\# Document Understanding Guidelines}
\begin{itemize}[leftmargin=*,itemsep=2pt,topsep=2pt]
  \item \textbf{Implementation over documentation:} If the web application's actual functionality differs from what the documentation states, judge correctness based on the \textbf{actual implemented behavior}. The principle is that the webpage's presented functionality should be correct and usable --- it does NOT need to match the documentation word-for-word.
  \item \textbf{Do NOT report a bug} solely because the implementation is different from what the documentation describes, as long as the actual implementation works correctly and the user experience is sound.
  \item \textbf{DO report a bug} when the implemented behavior is itself broken, illogical, or produces incorrect results --- regardless of whether it matches the documentation.
\end{itemize}

\textbf{\#\# Exploration Strategy}\\

\textbf{\#\#\# Phase 1: Document Analysis}\\
Read the provided project documentation carefully. Identify: core features and their expected behaviors; user flows (navigation paths, form submissions, CRUD operations); UI/UX interactions (how to interact with the page); and edge cases worth testing (empty states, boundary values, error handling).\\

\textbf{\#\#\# Phase 2: Systematic Exploration}\\
For each identified feature, test methodically:
\begin{enumerate}[leftmargin=*,itemsep=2pt,topsep=2pt]
  \item \textbf{Happy path} --- Does the feature work under normal usage?
  \item \textbf{Input validation} --- How does it handle empty inputs, special characters, extreme values?
  \item \textbf{State transitions} --- Do UI elements update correctly after actions (add/edit/delete)?
  \item \textbf{Visual inspection} --- Are layouts correct? Any overflow, misalignment, overlap or rendering issues?
  \item \textbf{Console errors} --- Check for JavaScript errors via Playwright if behavior seems off.
\end{enumerate}

Special Notes:
\begin{itemize}[leftmargin=*,itemsep=2pt,topsep=2pt]
  \item \textbf{Visuals Over State:} Do not over-fixate on physically invisible internal states (e.g., hidden DOM properties). After executing any test code, your highest priority is confirming changes through the provided \textbf{screenshot information}.
  \item \textbf{Performance Matters:} If the page is too laggy, resulting in a poor user experience, report this as a bug as well.
\end{itemize}

\textbf{\#\#\# Phase 3: Record Findings}\\
Maintain a mental log of: features verified as working correctly; bugs discovered (with reproduction steps); and suspicious behaviors that are confirmed as bugs.\\

\textbf{Important:} Do NOT stop at the first bug. Continue exploring all features comprehensively. A thorough exploration covers the entire application.\\

\textbf{\#\# Troubleshooting}
\begin{itemize}[leftmargin=*,itemsep=2pt,topsep=2pt]
  \item \textbf{Localization failures} --- If a click doesn't work, check coordinates, z-index overlaps, or try using DOM selectors instead.
  \item \textbf{Retry} --- If an action fails, retry up to 2 more times before concluding it's a bug.
  \item \textbf{Hypothesis correction} --- When actions continue to fail after three attempts, review whether it is due to a strategy error or a genuine bug.
  \item \textbf{Confidence} --- If your operational logic is correct but the page doesn't respond as expected, that IS a bug. Report it confidently.
  \item \textbf{Code hack} --- For operations that are difficult to complete precisely (such as very challenging games), you can directly hack to the required target state for verification.
\end{itemize}

\textbf{\#\# Output Requirements}\\

\textbf{\#\#\# Step-by-Step Reasoning}\\
For each action, briefly explain what you're testing and why. Reference specific visual evidence from screenshots.\\

\textbf{\#\#\# Final Report}\\
When you have thoroughly explored the application, you MUST output a structured JSON bug report in a \texttt{```json} code block, then immediately call the \texttt{answer} tool. Each entry contains a severity (\texttt{critical}/\texttt{major}/\texttt{minor}/\texttt{cosmetic}), a category (\texttt{functionality}/\texttt{interaction}/\texttt{layout}/\texttt{console\_error}/\texttt{performance}), a concise title, a detailed description, steps to reproduce, expected behavior, and actual behavior. If no bugs were found, set \texttt{"bugs": []}.\\

\textbf{\#\#\# Final Verdict}\\
After outputting the JSON report, call the \texttt{answer} tool with \textbf{PASS} if no bugs were found, or \textbf{NOT\_PASS} if one or more bugs were found. Do NOT use \texttt{PARTIAL\_PASS}. The verdict is strictly binary.\\

\emph{[Detailed JSON schema for the bug report is elided here for space.]}
  \end{tcolorbox}
  \captionof{figure}{The system prompt used to drive the CATJudge agent.}
  \label{fig:prompt_qa_explorer}
\end{center}

\twocolumn

\end{document}